\documentclass[journal,draftcls,onecolumn,12pt,twoside]{IEEEtran}
\usepackage{algorithm} %ctan.org\pkg\algorithms
\usepackage{algpseudocode}
\usepackage{graphicx}
\usepackage{amssymb}
\usepackage{amsfonts}
\usepackage{amsmath}
\usepackage{float}
\usepackage{url}
\usepackage{bm}
\usepackage[colorlinks=true]{hyperref}
\hypersetup{urlcolor=blue, citecolor=blue}

\IEEEoverridecommandlockouts
\newtheorem{Theorem}{Theorem}
\newtheorem{Proposition}{Proposition}
\newtheorem{Corollary}{Corollary}
\newtheorem{Definition}{Definition}
\newtheorem{Example}{Example}

\newtheorem{lemma}{Lemma}

\newenvironment{prof}{\textit{Proof\,:}} { $\square$}

\ifCLASSINFOpdf
\else
\fi
\begin{document}
\title{Lower bounds  on the lifting degree of single-edge and multiple-edge QC-LDPC codes by difference matrices}
%-----------------------------------------------------------main affiliation
%\author{
%\IEEEauthorblockN{Hassan~Khodaiemehr}
%\IEEEauthorblockA{Department of Mathematics and Computer \\ Science,
%Amirkabir University of Technology\\
% E-mail: h.khodaiemehr@aut.ac.ir}
% \and
%\IEEEauthorblockN{Mohammad-Reza~Sadeghi}
%\IEEEauthorblockA{Department of Mathematics and Computer \\ Science,
%Amirkabir University of Technology\\
%Email: msadeghi@aut.ac.ir}
%\and
%\IEEEauthorblockN{Amin~Sakzad}
%\IEEEauthorblockA{Department of ECSE, Monash\\
%University, Victoria, Australia\\
%E-mail: amin.sakzad@monash.edu}
%}
%--------------------------------------------------------------
\author{Farzane Amirzade and Mohammad-Reza~Sadeghi \\
\thanks{%
Manuscript received May ??, ????; revised November ??, ????.}
\thanks{  M.-R. Sadeghi is with the Department of Mathematics and Computer Science, Amirkabir University of Technology and F. Amirzade is with the Department of Mathematics, Shahrood University of Technology

(e-mail:  msadeghi@aut.ac.ir, famirzade@gmail.com).}
 \thanks{%
 Digital Object Identifier ????/TCOMM.?????}}

%\and
%\IEEEauthorblockN{Emanuele~Viterbo}
%\IEEEauthorblockA{Dept of ECSE, Monash University, Australia\\
%E-mail: emanuele.viterbo@monash.edu}}

%\author{Amin~Sakzad, and Mohammad-Reza~Sadeghi\thanks{A. Sakzad is with the ECSE Department at Monash University, Melbourne, Australia. M. R. Sadeghi is with the Faculty of Mathematics and Computer Science, Amirkabir University of Technology, Tehran, Iran.
%E-mails: amin.sakzad@monash.edu and msadeghi@aut.ac.ir.}}
\maketitle
\begin{abstract}
In this paper, we  define two matrices named as ``difference matrices", denoted by $D$ and $DD$ which significantly contribute to achieve  regular single-edge QC-LDPC codes with the shortest length and the certain girth as well as regular and irregular multiple-edge QC-LDPC codes.

 Making use of these matrices, we obtain necessary and sufficient conditions  to have single-edge $(m,n)$-regular QC-LDPC codes with girth 6 by which we achieve all non-isomorphic codes with  the minimum lifting degree, $N$, for $m=4$ and $5\leq n\leq 11$, and present an exponent matrix for each minimum distance. We attain the necessary and sufficient conditions to have a Tanner graph with girth 10. In this case we  also provide a lower bound on the lifting degree which is
 tighter than the existing bound. More important, for an exponent matrix whose first row and first column are all-zero, we demonstrate that the non-existence of 8-cycles proves the non-existence of 6-cycles  related to the first row of the exponent matrix too. All non-isomorphic QC-LDPC codes with girth 10 and $n=5,6$ whose numbers are more than those presented in the literature are provided. For $n=7,8$ we decrease the lifting degrees from 159 and 219 to 145 and 211, repectively. Moreover, necessary and sufficient conditions to have $(m,n)$-regular QC-LDPC codes with girth 12 as well as a lower bound on the lifting degree are achieved.
 
   For multiple-edge category, for the first time a lower bound on the lifting degree for both regular and irregular  QC-LDPC codes with girth 6 is achieved. We  analytically prove that if $\vec{B}_{ij}$ is $ij$-th element of an $m\times n$ exponent matrix $B$ then by taking three values  $A=\max\{2\sum_{j=1}^{n}{|\vec{B}_{ij}|\choose2};i=1,2\dots,m\}$, $B=\max\{2\sum_{i=1}^{m}{|\vec{B}_{ij}|\choose2};j=1,2\dots,n\}$ and $C=\max\{\sum_{j=1}^{n}|\vec{B}_{ij}|\times|\vec{B}_{i'j}|;i\neq i'; i,i'\in\{1,2\dots,m\}\}$  the minimum lifting degree is at least $N=\max\{A,B,C\}$. We also demonstrate that the achieved lower bounds on multiple-edge $(4,n)$-regular QC-LDPC codes with girth 6 are tight and the resultant codes have shorter length compared to their counterparts in single-edge codes. Additionaly difference matrices help to reduce the conditions of considering 6-cycles  from seven states to five states. We obtain  multiple-edge $(4,n)$-regular QC-LDPC codes with girth 8 and $n=4,6,8$ with the shortest length.

  \end{abstract}

% \fi
\IEEEpeerreviewmaketitle
\section{Introduction}
\IEEEPARstart{Q}  uasi-cyclic low-density parity-check codes (QC-LDPC codes) are an essential  category of LDPC codes that are preferred to other types of LDPC codes because of their favourably  practical and simple implementations. One of the most important representation of codes is Tanner graph in which the length of the shortest cycles, girth, has been known to influence the code performance. Since  Tanner graphs with short cycles do not produce good results, lots of efforts have been put into designing QC-LDPC codes with large girth.

There are two structures for constructing LDPC codes, algebraic-based and graph-theoretic-based. Of the latter the most well-known methods are progressive edge growth (PEG) and  protograph-based methods. There is a large body of work devoted to the protograph-based QC-LDPC code structures, some of which can be found in the reference section. Protograph itself can be divided into two categories, single-edge and multiple-edge protographs. They are denoted by a base matrix, $B_{ptg}$, whose elements identify the type of the protograph. If all the entries of $B_{ptg}$ are ones or zeros and the ones are substituted by $N\times N$ circulant permutation matrices (CPMs) and zero elements by an $N\times N$ zero matrix ($ZM$) then the null space of the resultant parity-check matrix constructs a QC-LDPC code and the corresponding $B_{ptg}$ is referred to as single-edge protograph. In $\cite{Multiple}$ it is shown that if $B_{ptg}$ contains an all-one submatrix of size $2\times3$ or $3\times2$ then Tanner graph has 12-cycles.  In QC-LDPC  codes, parameter $N$ stands for the lifting degree.  It is clear that the less the lifting degree is the shorter length QC-LDPC code is obtained. In order to construct a parity-check matrix with a given lifting degree $N$ all of 1-components of $B_{ptg}$ are replaced by $i;\ 0\leq i\leq N-1$. The obtained matrix is known as an exponent matrix and replacing all of non-zero elements and zeros by CPMs or ZMs provides us with the parity-check matrix.  In $\cite{2016}$, the authors claim to achieve all non-isomorphic fully connected single-edge $(3,n)$-regular QC-LDPC codes with the minimum lifting degree  for $n=4,\dots,12$ and girth 6, for $n=4,\dots,9$ and girth 8, $n=4,5,6$ and girth 10 and for  $n=4,5$ and girth 12. To obtain the results they make use of an exhaustive search and apply the well-known Fossorier's Lemma. There are also results for $m=3,4,5,6$ in $\cite{2004},\cite{2006},\cite{9},\cite{13},\cite{12},\cite{15}$ and different girths.  The lower bounds obtained in $\cite{15}$ on the lifting degree of $(m,n)$ regular QC-LDPC codes whose Tanner graphs have girth 6, 8 and 10 are $n$, $(m-1)(n-1)+1$ and $n(n-1)(m-1)+1$, respectively.

In this paper we define two matrices named as difference matrices, $D$, and  $DD$, from an exponent matrix. One of the advantages of providing these matrices is a  reduction in the complexity of the search algorithm to obtain the minimum lifting degree that achieves a certain girth. More importantly, it helps us to provide the necessary and sufficient condition for the difference matrix of a QC-LDPC code to have a Tanner graph with girth 6. By making use of these matrices we obtain all non-isomorphic fully  connected single-edge (4,n) QC-LDPC codes with the shortest length for $5\leq n\leq 11$. We also achieve their minimum distances and for each minimum distance we present an exponent matrix.  In addition, we investigate the conditions of difference matrices to have a Tanner graph with girth 8. The obtained conditions contribute us to prove the fact that if elements of the first row and the first column of the exponent matrix are zeros then the non-existence of 8-cycles guarantees the non-existence of 6-cycles related to the first row. This result expresses that if Tanner graph is free of 8-cycles then in order to have a Tanner graph with girth 10 there is no need to check $3\times3$ submatrices that contain the first all-zero row for considering the existence of 6-cycles. This consequence  significantly reduces the complexity of search. We apply the necessary and sufficient conditions of difference matrices on $(3,n)$ exponent matrices to have   QC-LDPC codes with girth 10, where $4\leq n\leq 8$. We conclude that the number of non-isomorphic codes obtained making use of  difference matrices for $n=5$ and $n=6$ is 4 and 2, respectively.  These numbers are more than those represnted in $\cite{2016}$, which are 3 and 1, respectively. However, in $\cite{2016}$ it is claimed to obtain all non-isompprphic exponent matrices by  an efficient exhaustive search. The minimum lifting degrees in $\cite{2006}$ for $n=7,8$ are 159 and 219, respectively, whereas applying difference matrices results in  $(3,n)$ QC-LDPC codes with girth 10 and shorter length, in these cases $N$ is 145 and 211, respectively.  We also provide analytical lower bounds for the lifting degree according to the girth. We demonstrate that for $(m,n)$  QC-LDPC codes whose Tanner graphs have girth 10 we have $N\geq {m\choose 2}n(n-1)+1$. Our proposed lower bound acknowledges those lifting degrees obtained search-based in $\cite{2016}$ for $m=3$ and $n=37,61,91$. Moreover our proposed lower bound improves the one in the literature. In  fact, our obtained lower bound is $\frac{m}{2}$ times as many as the one in $\cite{15}$, which is $n(n-1)(m-1)+1$.  We also consider all conditions of difference matrices to obtain  $(m,n)$  QC-LDPC codes with girth 12 and present a lower bound on the lifting degree. We prove that if $A$ is a set consisting of the values obtained from the left side of equation in Fossorier's Lemma for 6-cycle considerations then $N\geq |A|+{m\choose 2}n(n-1)+1$. 

Taking benefit of difference matrices we also achieve some results regarding to multiple-edge QC-LDPC codes for both regular and irregular categories. We summarize them in the following. According to our definition of difference matrices we prove that if $B$ is the exponent matrix of a multiple-edge QC-LDPC code of size $m\times n$ with girth 6 then by taking three variables,  $A=\max\{2\sum_{j=1}^{n}{|\vec{B}_{ij}|\choose2};i=1,2\dots,m\}$, $B=\max\{2\sum_{i=1}^{m}{|\vec{B}_{ij}|\choose2};j=1,2\dots,n\}$ and $C=\max\{\sum_{j=1}^{n}|\vec{B}_{ij}|\times|\vec{B}_{i'j}|;i\neq i'; i,i'\in\{1,2\dots,m\}\}$ we show that the minimum lifting degree is at least $N=\max\{A,B,C\}$. By applying the result on $(2,n)$ exponent matrices for which $|\vec{B}_{ij}|=2$ we have $N\geq 4n$. For $n=2,3,4,5$ and 6 we demonstrate the obtained lower bounds are tight. Furthermore, in comparison with single-edge $(4,2n)$ QC-LDPC codes with girth 6, the proposed multiple-edge QC-LDPC codes with girth 6 has shorter length. In order to obtain a multiple-edge QC-LDPC code with girth 8, seven types of submatrices have to be investigated. Thanks to difference matrices we decrease the investigation into five types and making use of them we obtain multiple-edge $(4,4)$, (4,6) and (4,8) QC-LDPC codes with the shortest length and  girth 8. 

The rest of the paper is organized as follows. Section II presents some basic notations. Section III includes four subsections to consider the equations related to $2k$-cycles, where $k=2,3,4,5$. Section IV is related to multiple-edge QC-LDPC codes in both regular and irregular types. In the last section we summarize our results.
\section{Preliminaries}\label{}

Let $N$ be an integer number. Consider the following exponent matrix $B=[\vec{B}_{ij}]$, where, $\vec{B}_{ij}=(b^{1}_{ij},b^{2}_{ij},\dots, b^{l}_{ij})$ in which $b^{r}_{ij}\in \lbrace 0,1,\cdots,N-1\rbrace$ and $\;b^{r}_{ij}\neq b^{r^\prime}_{ij}$ for $\;1\leq r<r^\prime\leq l,\;l\in \mathrm{N}$. 
\begingroup\fontsize{8.5pt}{11pt}\begin{align}\label{Rela2}
	B=\left[\begin{array}{cccc}
		\vec{B}_{00}&\vec{B}_{01}&\cdots &\vec{B}_{0(n-1)}\\
		\vec{B}_{10}&\vec{B}_{11}&\cdots &\vec{B}_{1(n-1)}\\
		\vdots &\vdots &\ddots &\vdots \\
		\vec{B}_{(m-1)0}&\vec{B}_{(m-1)1}&\cdots &\vec{B}_{(m-1)(n-1)}\\
		\end{array}\right],
		\end{align}\endgroup
		
		The $ij$-th element of the matrix $B$ is substituted by an $N\times N$ matrix $H_{ij}$ which is obtained by
		$$ H_{ij}=I^{b^{1}_{ij}}+I^{b^{2}_{ij}}+\cdots +I^{b^{l}_{ij}}.$$ 
		$I^{b^{r}_{ij}},\ 1\leq r\leq l$ is a circulant permutation matrix(CPM)  whose top row has a 1-component in the $({b^{r}_{ij}}+1)$-th column and  the others are zeros. The $i-$th row of the matrix of $I^{b^{r}_{ij}}$ is formed by  $i$ right cyclic shifts of the first row and clearly the first row is a right cyclic shift of the last row. If $\vec{B}_{ij}=(\infty)$ then it is replaced by a zero matrix. The null space of the parity-check matrix, $H$, which is presented in the following, provides us with a QC-LDPC code.  Note that, our definition of QC-LDPC codes covers both types of {\em single-edge protograph codes}  and {\em multiple-edge protograph codes}  that are produced by lifting each edge of their corresponding base graphs.
		
		\begingroup\fontsize{8.5pt}{11pt}\begin{align}\label{Rela1}
			H=\left[\begin{array}{cccc}
				H_{00}&H_{01}&\cdots &H_{0(n-1)}\\
				H_{10}&H_{11}&\cdots &H_{1(n-1)}\\
				\vdots &\vdots &\ddots &\vdots \\
				H_{(m-1)0}&H_{(m-1)1}&\cdots &H_{(m-1)(n-1)}\\
				\end{array}\right],
				\end{align}\endgroup
In single-edge protographs whose base matrices have elements of cardinality of at most 1, the necessary and sufficient condition for the existence of cycles of length $2k$ is provided in $\cite{2004}$. This well-known result is our principle tool and we summarise it as follows.

If
\vspace*{-0.16cm}
\begingroup\fontsize{8.5pt}{11pt}  \begin{equation}\label{Rela3}
\sum_{i=0}^{k-1}(b_{m_in_i}-b_{m_in_{i+1}})=0  \mod N,
\end{equation}\endgroup
where $n_k=n_0,\ m_i\neq m_{i+1},\ n_i\neq n_{i+1}$ and $b_{m_in_i}$ is the $(m_i,n_i)$-th entry of $B$ then the Tanner graph of the parity-check matrix has $2k$-cycles.
To compute the number of 4-cycles all of $2\times2$ submatrices have to be investigated. And for 6-cycles all of $3\times3$ submatrices are considered. But, computing Equation (3) is time-consuming especially when $2k$ is more than 6. For example, in order to investigate the existence or non-existence of 8-cycles we have to consider all of submatrices of sizes $2\times2,\ 2\times3,\ 2\times4,\ 3\times2,\ 3\times3,\ 3\times4,\ 4\times2,\ 4\times3$ and  $4\times4$.  In this paper our goal is to reduce the conditions on which the existence of $2k$-cycles are considered. To reach our goal we define matrices named as {\em difference matrices} which are denoted by $D$ and $DD$.

It is clear that an equivalent version of Equation (3) is: \vspace*{-0.16cm}
\begingroup\fontsize{8.5pt}{11pt}  \begin{equation}\label{Rela4}
\sum_{i=0}^{k^{\prime}-1} \left( B^{r^{}_i}_{m_{i}n_{i}} - B^{r^{\prime}_i}_{m_{i}n_{i+1}} \right) = 0  \mod N , 
\end{equation}
\endgroup
where $n_{k^{\prime}}=n_{0}$, $r^{}_{i}\neq r^{\prime}_{i}$ if $n_{i} = n_{i+1}$, $ r^{\prime}_{i}\neq r^{}_{i+1}$ if $m_{i} =m_{i+1}$, $B^{r^{}_i}_{m_{i}n_{i}}$ is the $r^{}_i$-th entry of $\vec{B}_{m_{i}n_{i}}$ and $\vec{B}_{m_{i}n_{i}}$ is the $(m_i n_i)$-th entry of $B$.	

\section{Single-edge protograph based QC-LDPC codes }\label{}
In this section we investigate Equation (3) for $2k$-cycles, especially those of sizes  4, 6, 8 and 10. And for each of these considerations we investigate the conditions of difference matrices $D$ and/or $DD$. We also provide lower bounds for the lifting degree, where the Tanner graph is free of the 8-cycles or 10-cycles.  Then we compare the complexity of search algorithm when we apply Fossorier's Lemma to the exponent matrix, $B$, with  when we apply the matrices, $D$ and $DD$. For each size of cycles we provide a subsection. But before, we explain how to construct two difference matrices $D$ and $DD$ as follows.
\begin{Definition}
	Suppose $B$ is an exponent matrix shown above, the difference matrix, $D$, is as follows:
	\begin{center}
		$D=\left[\begin{array}{lllll}
		b_{00}-b_{10} & \dots & b_{0(n-1)}-b_{1(n-1)}\\
		\vdots    &  \ddots & \vdots\\
		b_{00}-b_{(m-1)0}  & \dots & b_{0(n-1)}-b_{(m-1)(n-1)}\\
		b_{10}-b_{20}  & \dots & b_{1(n-1)-b_{2(n-1)}}\\
		\vdots      &  \ddots & \vdots\\
		b_{10}-b_{(m-1)0}  & \dots & b_{1(n-1)}-b_{(m-1)(n-1)}\\
		\vdots      & \ddots & \vdots\\
		b_{(m-2)0}-b_{(m-1)0} & \dots & b_{(m-2)(n-1)}-b_{(m-1)(n-1)}\\
		\end{array}\right].$
	\end{center}
	If $B$ includes an $(\infty)$ as its element then we put $\infty-b_{ij}=\infty$ and $\infty-\infty=\infty$.
\end{Definition}

We also utilize another matrix to reduce the complexity. It is obtained from $D$, and we denote it by $DD$.
\begin{Definition}
	Suppose $D_{ij}$ and $D_{ij'}$ are two elements of $i$-th row of the difference matrix, $D$, then $(D_{ij}-D_{ij'},D_{ij'}-D_{ij})$, of course $\mod\ N$, is an element of $i$-th row of $DD$. So $DD$ is a matrix of size ${m\choose 2}\times{n\choose 2}$.
\end{Definition}
\begin{Example}
	we  obtain the difference matrix $D$ from a $3\times4$ exponent matrix with the lifting degree $N=37$,
	\begin{center}
		$B=\left[\begin{array}{cccc}
		0 & 0 & 0 & 0 \\
		0 & 1 & 3 & 24 \\
		0 & 27 & 7 & 19 \\
		\end{array}\right]$, $D=\left[\begin{array}{cccc}
		0 & -1 & -3 & -24\\
		0 & -27 & -7 & -19\\
		0 & -26& -4 & 5 \\
		\end{array}\right]$.
	\end{center}
	The $3\times6$ matrix, $DD$, is presented as follows:
	\begin{center}
		$\left[\begin{array}{cccccc}
		(1,36) & (3,34) & (13,24) & (2,35) & (23,14) & (21,16)\\
		(10,27) & (30,7) & (18,19) & (17,20) & (29,8) & (12,25) \\
		(11,26) & (33,4) & (5,32) & (22,15) & (6,31) & (28,9) \\
		\end{array}\right]$.
	\end{center}
\end{Example}
\subsection{4-cycles}\label{}
In order to consider 4-cycles, Equation (3) is investigated for every $2\times2$ submatrix. Suppose a $2\times2$ submatrix in two rows $i_1$ and $i_2$ and two columns $j_1$ and $j_2$ is under consideration, then Equation (3) gives $(b_{i_1j_1}-b_{i_1j_2})+(b_{i_2j_2}-b_{i_2j_1})=0\ \ (\mod\ N)$. But in order to consider 4-cycles making use of the difference matrix, $D$, we rearrange the latter equality as follows:
\begin{center}
	$(b_{i_1j_1}-b_{i_2j_1})-(b_{i_1j_2}-b_{i_2j_2})=0\ \ (\mod\ N).$
\end{center}

\begin{Proposition}
The Tanner graph is 4-cycle free if and only if there are no equal elements in any row of the difference matrix, $D$.
\end{Proposition}
\begin{prof}
Given $1\leq i_{1},i_{2} \leq m\;(i_{1} < i_{2})$. The expression $(b_{i_1j_1}-b_{i_2j_1})$ is an element of $D$ located in $i$-th row and $j_1$-th column, where $i={{m}\choose{2}}-{{m-i_{1}+1}\choose{2}ý}+i_{2}-i_{1}$, and $(b_{i_1j_2}-b_{i_2j_2})$ is another element of $D$ in $i$-th row and $j_2$-th column. So we conclude that if $b_{i_1j_1}-b_{i_2j_1}=b_{i_1j_2}-b_{i_2j_2} \ \ (\mod\ N)$ then the  Tanner graph has 4-cycles.
\end{prof}
\begin{Corollary}
	The Tanner graph is 4-cycle free if and only if there is no zero element in the difference matrix, $DD$.
	\end{Corollary}
In $\cite{2016}$ all non-isomorphich $(3,n)$-regular single-edge QC-LDPC codes for $4\leq n\leq12$ along with their minimum distances are represented. In order to extend the result we make use of difference matrices and provide the number of non-isomorphic (or simply NI numbers) $(4,n)$-regular single-edge QC-LDPC codes for $5\leq n\leq11$ and present one exponent matrix for each of the existing minimum distance in the following table. The first row and column of the exponent matrices are all-zero which are omitted.

\begin{table}[h]
	\begin{center}
		\begin{tabular}{|c|c|c|c|c|}
			\hline
			Column weight & $d_c=5$ & $d_c=6$ & $d_c=7$& $d_c=8$\\
			\hline
			Minimum distance & $d_{min}=8$ & $d_{min}=8,10$  & $d_{min}=8,10$& $d_{min}=6,8,10$\\
			\hline
			 Lifting degree & $N=5$ &  $N=7$  & $N=7$ & $N=10$ \\
			\hline
			 NI numbers&1 & 2 & 2& 2996\\
			\hline
			B with $d_{min}=10$&- &  $\begin{array}{cccccc}
			1 & 3 & 4 & 5 & 6 \\
			5 & 1 & 6 & 4 & 2 \\
			2 & 6 & 1 & 3 & 5
			\end{array}$ &  $\begin{array}{cccccc}
			1 & 2 & 3 & 4 & 5 & 6 \\
			2 &4 & 6 & 1 & 3 & 5 \\
			3 & 6 & 2 & 5 & 1 & 4
			\end{array}$ & $\begin{array}{cccccccc}
			1 & 2 & 3 & 4 & 5 & 6 & 7 \\
			2 & 1 & 5 & 7 & 9 & 4 & 3 \\
			5 & 7& 2 & 6 & 1 & 9 & 4
			\end{array}$\\
			\hline
			B  with $d_{min}=8$&$\begin{array}{ccccc}
				1 & 2 & 3 & 4 \\
				3 & 1 & 4 & 2 \\
				2 & 4 & 1 & 3
			\end{array}$ & $\begin{array}{ccccc}
			 1 & 3 & 4 & 5 & 6 \\
			 5 & 1 & 6 & 4 & 2 \\
			 3 & 2 & 5 & 1 & 4 
			 \end{array}$ & $\begin{array}{cccccc}
			  1 & 2 & 3 & 4 & 5 & 6 \\
			  2 &4 & 6 & 1 & 3 & 5 \\
			  4 & 1 & 5 & 2 & 6 & 3  
			 \end{array}$& $\begin{array}{cccccccc}
			 1 & 2 & 3 & 4 & 5 & 6 & 7 \\
			 2 & 1 & 5 & 7 & 9 & 4 & 3 \\
			 3 & 6 & 2 & 1 & 8 & 7 & 5
			 \end{array}$\\
			 \hline
			 B with $d_{min}=6$&- & -& -&$\begin{array}{cccccccc}
			 1 & 2 & 3 & 4 & 5 & 6 & 7 \\
			 2 & 1 & 5 & 8 & 3 & 9 & 4 \\
			 7 & 3 & 6& 1 & 9 & 8& 2
			 \end{array}$\\
			 \hline
		\end{tabular}
		\caption{One of Non-isomorphic $(4,n)$-regular single-edge QC-LDPC codes with the existing minimum distances and column weight value 5, 6, 7 and 8}
	\end{center}
\end{table}
\begin{table}[h]
	\begin{center}
		\begin{tabular}{|c|c|c|c|}
			\hline
			Column weight & $d_c=9$& $d_c=10$&$d_c=11$\\
			\hline
			Minimum distance & $d_{min}=6,8,10$ & $d_{min}=8,10$  & $d_{min}=8,10$\\
			\hline
			Lifting degree & $N=10$ &  $N=11$  & $N=11$ \\
			\hline
			NI numbers&29 & 24 & 7\\
			\hline
			B with $d_{min}=10$ & $\begin{array}{ccccccccc}
			1 & 2 & 3 & 4 & 5 & 6 & 7 & 8 \\
			2 & 5 & 7 & 9 & 4 & 8 & 3 & 6 \\
			5 & 3 & 9 & 6 & 8 & 4 & 2 & 7
			\end{array}$ &  $\begin{array}{ccccccccc}
			1 & 2 & 3 & 4 & 5 & 6 & 7 & 9&10 \\
			10& 9& 8 & 7 & 6 & 5 & 4 & 2&1\\
			2 & 4 & 6 & 8 & 10& 1 & 3 & 7&9
			\end{array}$ &  $\begin{array}{ccccccccccc}
			1 & 2 & 3 & 4 & 5 & 6 & 7 & 8 & 9 & 10\\
			7 & 3 & 10 & 6 & 2 & 9 & 5 & 1 & 8 & 4\\
			2 & 4 & 6 & 8 & 10& 1 & 3 & 5 & 7 & 9  
			\end{array}$\\
			\hline
		B with $d_{min}=8$& $\begin{array}{cccccccc}
		1 & 2 & 3 & 4 & 5 & 6 & 7 & 8 \\
		2 & 1 & 6 & 8& 7 & 3 & 5 & 4\\
		3 & 6 & 8 & 5& 1 & 9 & 4 & 7
		\end{array}$ & $\begin{array}{ccccccccc}
			1 & 2 & 3 & 4 & 5 & 6 & 7 & 9&10 \\
			5 & 10 & 4 & 2 & 7 & 9 & 1 & 8&6\\
			4 & 8 & 1 & 5 & 9 & 2 & 6 & 3&7
			\end{array}$ & $\begin{array}{ccccccccccc}
			1 & 2 & 3 & 4 & 5 & 6 & 7 & 8 & 9 & 10\\
			2 & 6 & 9& 7 & 4 & 3 & 1 & 10&5 & 8\\
			3 & 1 & 7 & 9 & 8&2 & 4 & 6 & 10& 5  
			\end{array}$\\
			\hline
			B with $d_{min}=6$& $\begin{array}{cccccccc}
			1 & 2 & 3 & 4 & 5 & 6 & 7 & 8 \\
			2 & 5 & 1 & 9& 7 & 3 & 6 & 4\\
			5 & 1& 9 & 6& 2 & 4 & 8 & 3
			\end{array}$ & -& -\\
				\hline
		\end{tabular}
		\caption{One of Non-isomorphic $(4,n)$-regular single-edge QC-LDPC codes with the existing minimum distances and column weight value 9, 10 and 11.}
	\end{center}
\end{table}
\subsection{6-cycles}\label{}
In this subsection we first consider 6-cycles for three types of exponent matrices with row weights 3, 4 and 5. Then we present a structure to consider $6$-cycles in $m\times n$ exponent matrices.

Suppose $B$ has three rows. It is clear that for each three columns Equation (3) has to be investigated. Suppose that  $B'$ and $D'$ are a $3\times3$  submatrix of $B$ and its corresponding  submatrix of $D$, respectively.
\begin{center}
$B'=\left[\begin{array}{ccccc}
b_{i_0j_0} & b_{i_0j_1} & b_{i_0j_2}\\
b_{i_1j_0} & b_{i_1j_1} & b_{i_1j_2}\\
b_{i_2j_0} & b_{i_2j_1} & b_{i_2j_2}\\
\end{array}\right]$, $D'=\left[\begin{array}{ccccc}
b_{i_0j_0}-b_{i_1j_0} & b_{i_0j_1}-b_{i_1j_1} & b_{i_0j_2}-b_{i_1j_2}\\
b_{i_0j_0}-b_{i_2j_0} & b_{i_0j_1}-b_{i_2j_1} & b_{i_0j_2}-b_{i_2j_2}\\
b_{i_1j_0}-b_{i_2j_0} & b_{i_1j_1}-b_{i_2j_1} & b_{i_1j_2}-b_{i_2j_2}\\
\end{array}\right]$
\end{center}

The left side of Equation (3) gives the following 6 equalities. Moreover, for each expression we obtain the corresponding equality making use of the elements of the difference matrix, $D$,

\begin{itemize}
\item $b_{i_0j_0}-b_{i_0j_1}+ b_{i_1j_1}-b_{i_1j_2}+b_{i_2j_2}-b_{i_2j_0}=(b_{i_0j_0}-b_{i_2j_0})-(b_{i_0j_1}-b_{i_1j_1})-(b_{i_1j_2}-b_{i_2j_2})$
\item $b_{i_0j_1}-b_{i_0j_0}+ b_{i_1j_0}-b_{i_1j_2}+b_{i_2j_2}-b_{i_2j_1}=-(b_{i_0j_0}-b_{i_1j_0})+(b_{i_0j_1}-b_{i_2j_1})-(b_{i_1j_2}-b_{i_2j_2})$
\item $b_{i_0j_0}-b_{i_0j_2}+ b_{i_1j_2}-b_{i_1j_1}+b_{i_2j_1}-b_{i_2j_0}=(b_{i_0j_0}-b_{i_2j_0})-(b_{i_1j_1}-b_{i_2j_1})-(b_{i_0j_2}-b_{i_1j_2})$
\item $b_{i_0j_2}-b_{i_0j_0}+ b_{i_1j_0}-b_{i_1j_1}+b_{i_2j_1}-b_{i_2j_2}=-(b_{i_0j_0}-b_{i_1j_0})-(b_{i_1j_1}-b_{i_2j_1})+(b_{i_0j_2}-b_{i_2j_2})$
\item $b_{i_0j_1}-b_{i_0j_2}+ b_{i_1j_2}-b_{i_1j_0}+b_{i_2j_0}-b_{i_2j_1}=-(b_{i_1j_0}-b_{i_2j_0})+(b_{i_0j_1}-b_{i_2j_1})-(b_{i_0j_2}-b_{i_1j_2})$
\item $b_{i_0j_2}-b_{i_0j_1}+ b_{i_1j_1}-b_{i_1j_0}+b_{i_2j_0}-b_{i_2j_2}=-(b_{i_1j_0}-b_{i_2j_0})-(b_{i_0j_1}-b_{i_1j_1})+(b_{i_0j_2}-b_{i_2j_2}).
$
\end{itemize}
As a result, each expression in the right side of the six equations above includes three elements of a distributed diagonal of $D'$ with the assumption that the elements in the first and the third rows are multiplied by $-1$. So we proved the following lemma.
\begin{lemma}
Let $D$ be a difference matrix corresponding to a $3\times n$ exponent matrix, $B$. Then the Tanner graph is 6-cycle free if and only if $-D_{1j_1}+D_{2j_2}-D_{3j_3}\neq0\ (mod\ N)$ for all $j_{1}\neq j_{2},j_{1}\neq j_{3},j_{2}\neq j_{3}\;(j_{1},j_{2},j_{3}\in \{ 0,1,\ldots,(n-1)\})$.
\end{lemma}
Now, we consider 6-cycles in a $4\times n$ exponent matrix. Take a $4\times3$ submatrix of $B$ and its corresponding submatrix of $D$ as follows:

\begin{center}
$B'=\left[\begin{array}{ccccc}
b_{i_0j_0} & b_{i_0j_1} & b_{i_0j_2}\\
b_{i_1j_0} & b_{i_1j_1} & b_{i_1j_2}\\
b_{i_2j_0} & b_{i_2j_1} & b_{i_2j_2}\\
b_{i_3j_0} & b_{i_3j_1} & b_{i_3j_2}\\
\end{array}\right]$, $D'=\left[\begin{array}{ccccc}
b_{i_0j_0}-b_{i_1j_0} & b_{i_0j_1}-b_{i_1j_1} & b_{i_0j_2}-b_{i_1j_2}\\
b_{i_0j_0}-b_{i_2j_0} & b_{i_0j_1}-b_{i_2j_1} & b_{i_0j_2}-b_{i_2j_2}\\
b_{i_0j_0}-b_{i_3j_0} & b_{i_0j_1}-b_{i_3j_1} & b_{i_0j_2}-b_{i_3j_2}\\
b_{i_1j_0}-b_{i_2j_0} & b_{i_1j_1}-b_{i_2j_1} & b_{i_1j_2}-b_{i_2j_2}\\
b_{i_1j_0}-b_{i_3j_0} & b_{i_1j_1}-b_{i_3j_1} & b_{i_1j_2}-b_{i_3j_2}\\
b_{i_2j_0}-b_{i_3j_0} & b_{i_2j_1}-b_{i_3j_1} & b_{i_2j_2}-b_{i_3j_2}\\
\end{array}\right]$.
\end{center}
The submatrix $B'$ contains four $3\times3$ submatrices. Suppose the first three rows are chosen to consider Equation (3). So the obtained equalities are as the same as 6 equalities mentioned above from which the first one is $b_{i_0j_0}-b_{i_0j_1}+ b_{i_1j_1}-b_{i_1j_2}+b_{i_2j_2}-b_{i_2j_0}=(b_{i_0j_0}-b_{i_2j_0})-(b_{i_0j_1}-b_{i_1j_1})-(b_{i_1j_2}-b_{i_2j_2})$. We conclude that $-(b_{i_0j_1}-b_{i_1j_1})+(b_{i_0j_0}-b_{i_2j_0})-(b_{i_1j_2}-b_{i_2j_2})=-D'_{12}+D'_{21}-D'_{43}$.
Investigating all equations in four submatrices results in the following lemma.
\begin{lemma}
Let $D$ be a difference matrix corresponding to a $4\times n$ exponent matrix, $B$. The Tanner graph is 6-cycle free if and only if for all $j_{1}\neq j_{2},j_{1}\neq j_{3},j_{2}\neq j_{3}\;(j_{1},j_{2},j_{3}\in \{ 0,1,\ldots,(n-1)\})$ we have:
\begin{itemize}
\item $1)-D_{1j_1}+D_{2j_2}-D_{4j_3}\neq0\ \ \ \ \ 2) -D_{1j_1}+D_{3j_2}-D_{5j_3}\neq0$,
\item $3)-D_{2j_1}+D_{3j_2}-D_{6j_3}\neq0 \ \ \ \ \ 4) -D_{4j_1}+D_{5j_2}-D_{6j_3}\neq0$,
\end{itemize}
All of the expressions are computed in modular $N$.
\end{lemma}
Similarly, we consider all of $3\times3$ submatrices of the difference matrix corresponding to a $5\times n$ exponent matrix. And the result is as the following lemma.
\begin{lemma}
Let $D$ be a difference matrix corresponding to a $5\times n$ exponent matrix, $B$. The Tanner graph is 6-cycle free if and only if for all $j_{1}\neq j_{2},j_{1}\neq j_{3},j_{2}\neq j_{3}\;(j_{1},j_{2},j_{3}\in \{ 0,1,\ldots,(n-1)\})$ we have:
\begin{itemize}
\item $1)-D_{1j_1}+D_{2j_2}-D_{5j_3}\neq0,\ \ \ \ \ \ \ \ \ 2)-D_{1j_1}+D_{3j_2}-D_{6j_3}\neq0$,
\item $3)-D_{1j_1}+D_{4j_2}-D_{7j_3}\neq0,\ \ \ \ \ \ \ \ \ 4)-D_{2j_1}+D_{3j_2}-D_{8j_3}\neq0$,
\item $5)-D_{2j_1}+D_{4j_2}-D_{9j_3}\neq0,\ \ \ \ \ \ \ \ \ 6)-D_{3j_1}+D_{4j_2}-D_{10j_3}\neq0$,
\item $7)-D_{5j_1}+D_{6j_2}-D_{8j_3}\neq0,\ \ \ \ \ \ \ \ \ 8)-D_{5j_1}+D_{7j_2}-D_{9j_3}\neq0$,
\item $9)-D_{6j_1}+D_{7j_2}-D_{10j_3}\neq0$ $\ \ \ \ \ \ \ \  10)-D_{8j_1}+D_{9j_2}-D_{10j_3}\neq0$.
\end{itemize}
All of the expressions are computed in modular $N$.
\end{lemma}
Generally, if 6-cycles of an $m\times n$ exponent matrix are under consideration then by obtaining the corresponding difference matrix, $D$, we have to determine the existence of 6-cycles for some triples of form $(i,j,k)$, where $i,j,k\in\{0,1,\dots,(_{2}^{m})\}$. For example, as mentioned in the lemmas above, triples for exponent matrices with four rows are (1,2,4), (1,3,5), (2,3,6) and (4,5,6). And triples for exponent matrices with five rows are (1,2,5), (1,3,6), (1,4,7), (2,3,8), (2,4,9), (3,4,10), (5,6,8), (5,7,9), (6,7,10) and (8,9,10). In order to recognize which rows form the mentioned triples we provide the following method. 

Let $D$ be a difference matrix corresponding to an $m\times n$ exponent matrix, $B$. It is clear that in a $3\times3$ submatrix, which provides 6-cycles, two elements of each row and column of the submatrix occur in Equation (3). So there are two elements with a row index $i$, two elements with a row index $i'$ and two elements with a row index $i''$. In other words, if $\pm(b_{ij}-b_{i'j})$ and $\pm(b_{ij'}-b_{i''j'})$ are two differences which occur in Equation (3) which are also elements of $i_1$-th row and $i_2$-th row of $D$, respectively, then the third row is the one which includes $b_{ij''}-b_{i''j''}$ or $-(b_{ij''}-b_{i''j''})$. If it occurs in $i_3$-th row of $D$ then the corresponding triple is $(i_1,i_2,i_3)$.

In $\cite{2016}$ 6-cycle considerations for $3\times n$ exponent matrices whose first row and column are all-zero are converted into three types. In the following Lemma we demonstrate that  making use of the difference matrix $DD$ it is sufficient to investigate just one type. Moreover, this type can be utilized for all $m\times n$ exponent matrices whose first row and column are all-zero.  In the following Lemma $DD_{ij}$ and $N-DD_{ij}$ are the first and second component of $ij$-th  element of $DD$, respectively.
\begin{lemma}
	Let $DD$ be a difference matrix corresponding to a $m\times n$ exponent matrix, $B$. Suppose the first row of the $3\times 3$ submatrix under 6-cycle consideration is all-zero. If Tanner graph is 6-cycle free then we have:
	
	\noindent $DD_{ij_2}\neq DD_{i'j_3},\ DD_{ij_2}\neq DD_{i'j_1},\ DD_{ij_1}\neq DD_{i'j_2},\ DD_{ij_1}\neq -DD_{i'j_3},\ DD_{ij_3}\neq -DD_{i'j_1}$ and $DD_{ij_3}\neq DD_{i'j_2}$,
	 
	  where $i\neq i',\ i,i'\in\{1,2,m-1\}$. Disjoint columns $j_{1},j_{2}$ and $j_{3}$ of $DD$ are corresponding to the three columns of $D$.
\end{lemma}
Taking benefit of the above Lemma, 6-cycle considerations for a $4\times n$ exponent matrix has two types. First type belongs to equations  1,2 and 3 in Lemma 2 which are investigated by the difference matrix, $DD$.  And the second one is $-D_{4j_1}+D_{5j_2}-D_{6j_3}\neq0$. In addition, for a $5\times n$ exponent matrix, equations 1 to 6 in Lemma 3 are investigated by the difference matrix, $DD$. So just the last four equations in Lemma 3 are considered by the difference matrix, $D$.  Besides declining  the complexity of search algorithm, Lemma 4 has significant  contribution which we will see in finding exponent matrices whose Tanner graph has girth 10. This merit will be proposed in the next 8-cycle consideration.
\subsection{8-cycles}\label{}
Nine types of submatrices  have to be investigated in order to obtain a QC-LDPC code avoiding 8-cycles. The mentioned submatrices are of sizes $2\times2,2\times3,2\times4,3\times2,3\times3,3\times4,4\times2,4\times3$ and $4\times4$. Taking benefits of two matrices $D$ and $DD$ we reduce the complexity of search space in a great extent. We prove that in order to construct an exponent matrix with girth 10 there is no need to  investigate all of the mentioned submatrices. More important, if the first row and column of the exponent matrix are all-zero then there is no need to consider 6-cycles on $3\times3$ submatrices whose first row are the all-zero. Because the non-existence of 8-cycles guarantees the non-existence of 6-cycles too. In this subsection we also provide a lower bound on the lifting degree of QC-LDPC codes whose Tanner graphs have girth at least 10. We prove that the minimum lifting degree of QC-LDPC codes with girth 10 is $2{n\choose 2}{m\choose 2}+1$. Our proposed lower bound  improves the one in $\cite{15}$, which is $n(n-1)(m-1)+1$.  

In the following Lemma we provide our principle tool to consider 8-cycles.
\begin{lemma}
If an 8-cycle contains more than two elements of a column of the exponent matrix then it contains two $2\times2$ submatrices.
\end{lemma}
\begin{prof}
We prove one of the types of submatrices which satisfies the condition of Lemma, others are alike and are omitted. Suppose the left side of  Equation (3) is achieved by a $4\times3$ submatrix. For example, it is $b_{i_0j_0}-b_{i_0j_1}+b_{i_1j_1}-b_{i_1j_2}+b_{i_2j_2}-b_{i_2j_1}+b_{i_3j_1}-b_{i_3j_0}$. The two $2\times2$ submatrices are \begin{center}
$B'=\left[\begin{array}{cc}
b_{i_0j_0} & b_{i_0j_1} \\
b_{i_3j_0} & b_{i_3j_1} \\
\end{array}\right]$ and $B''=\left[\begin{array}{cc}
b_{i_1j_1} & b_{i_1j_2} \\
b_{i_2j_1} & b_{i_2j_2} \\
\end{array}\right]$.
\end{center}
\end{prof}
The advantage of the above lemma is that we can obtain an equivalence for Equation (3) whose elements belong to two rows of the difference matrix. Whereas, the left side of Equation (3) can be rearranged in a way that more than two rows of the difference matrix are involved. For example, $b_{i_0j_0}-b_{i_0j_1}+b_{i_1j_1}-b_{i_1j_2}+b_{i_2j_2}-b_{i_2j_1}+b_{i_3j_1}-b_{i_3j_0}=b_{i_0j_0}-b_{i_3j_0}-(b_{i_0j_1}-b_{i_1j_1})-(b_{i_1j_2}-b_{i_2j_2})-(b_{i_2j_1}-b_{i_3j_1})=D_{i'j_0}-D_{ij_1}-D_{i''j_2}-D_{i'''j_1}$ which contains four rows of the difference matrix, where $i<i'<i''<i'''$.

In the following theorem we make clear the advantage of applying the matrix, $DD$, in our 8-cycle considerations. The proof of the theorem is provided in Appendix A.
\begin{Theorem}
A QC-LDPC code has no 8-cycles if and only if:
\begin{itemize}
\item the matrix obtained by multiplying the difference matrix, $DD$, by 2 (or simply $2DD$) has no zero element and,
\item the matrix $DD$ has no repeated elements and,
\item each $4\times4$ submatrix of the difference matrix, $D$, satisfies the following inequality
\begin{center}
$D_{ij_0}-D_{i'j_1}-D_{i''j_2}-D_{i'''j_3}\neq0\ (\mod\ N),$
\end{center}
where $i,i',i'',i'''\in\{0,1,\dots,{m\choose 2}\}$ and $j_0,j_1,j_2,j_3\in\{0,1,\dots,n-1\}$.
\end{itemize}
\end{Theorem}

According to theorem above, if we use the matrix $DD$ then there is no need to consider all of submatrices of sizes $2\times3,\ 2\times4,\ 3\times2,\ 3\times3,\ 3\times4,\ 4\times2$ and $4\times3$, but instead, we consider the non-existence of repeated elements in the matrix, $DD$.
\begin{Example}
	Suppose $B$ is the $3\times4$ exponent matrix given in this section. As $N=37$ is an odd number, for $ij$-th  element of $2DD$ we have $2DD_{ij}\neq0\ (mod\ N)$. So $B$ holds in the first condition of the above theorem. As we see, in this case no computation is required. By comparing  each two elements of the matrix, $DD$, we conclude that there are no repeated elements in $DD$. Hence, $B$ holds in the second condition of the above theorem. Moreover, since the last condition of the above theorem needs 4 rows it is not considered for the given exponent matrix. As a whole, the Tanner graph is 8-cycle free. Whereas, the number of Equations (3) for submatrices of sizes $2\times2,\ 2\times3,\ 2\times4,\ 3\times2,\ 3\times3$ and  $3\times4$ is 168. 
\end{Example}
In appendix A, we compute the number of equations required to consider 8-cycles when we use Fossorier's equation. Note that, the number of computations for each type of submatrices is nine times as many as the number of equations. As an example, in the table III we provide the number of equations, when we use Fossorier's Lemma, for $m=3,4,5,6$ as follows.
\begin{table}[h]
	\begin{center}
		\begin{tabular}{|c|c|}
			\hline
			& The number of Fossorier's equations required to consider \\
			\hline
			$m=3$ &  $6{n\choose 2}+27{n\choose 3}+24{n\choose 4}$\\ 
			\hline    
			$m=4$  & $24{n\choose 2}+102{n\choose 3}+68{n\choose 4}$ \\
			\hline
			$m=5$ & $70{n\choose 2}+270{n\choose 3}+160{n\choose 4}$ \\
			\hline
			$m=6$ & $165{n\choose 2}+585{n\choose 3}+330{n\choose 4}$ \\
			\hline
		\end{tabular}
		\caption{}
	\end{center}
\end{table}
\begin{Corollary}
	Let $B$ be an exponent matrix whose Tanner graph is 4-cycle free. The number of computations required to obtain a QC-LDPC code whose Tanner graph is 8-cycle free too when we make use of Equation (3) is 
	\begin{center}
		$9{n\choose 2}{m\choose 2}({n-2\choose 2}(\frac{2}{9}{m-2\choose 2}+3m-4)+{n-2\choose 1}(\frac{2}{3}{m-2\choose 2}+2m-3)+m)$.
	\end{center}
	And if we utilize the matrix $DD$, given $N$ is an odd number, then the number of computations is 
	\begin{center}
		$n{m\choose 2}+3{n\choose 2}{m\choose 2}+{{{n\choose 2}{m\choose 2}}\choose 2}+5{n\choose 4}{m\choose 4}$.
	\end{center}
	And if $N$ is an even number then the number of computation is 
		\begin{center}
			$n{m\choose 2}+5{n\choose 2}{m\choose 2}+{{{n\choose 2}{m\choose 2}}\choose 2}+5{n\choose 4}{m\choose 4}$.
		\end{center}
\end{Corollary}
Note that, in the above corollary the required computations to construct the matrices $D$ and $DD$ are also taken into account, which are $n{m\choose 2}$ and $3{n\choose 2}{m\choose 2}$, respectively. Moreover in order to demonstarte the non-existence of repeated elements in $DD$, it is sufficient to consider the smaller element of each pair appeared in $DD$. If they provide a subset of $1,2,\dots,[\frac{N}{2}]$ of the length ${n\choose 2}{m\choose 2}$ then the exponent matrix is 8-cycle free. The number of computations in this case is ${{{n\choose 2}{m\choose 2}}\choose 2}$. More imprtantly, in the following we demonstrate the significant reduction in the search space by making use of our proposed difference matrices.
\begin{Corollary}
Suppose $B$ is a $3\times n$ exponent matrix. If the first row and column of $B$ are all-zero then the non-existence of 8-cycles guarantees the non-existence of 6 cycles.
\end{Corollary} 
\begin{prof}
 If Tanner graph is 8-cycle free then according to Theorem 1  the difference matrix, $DD$, has no repeated elements. Therefore, all of inequalities in lemma 4 occur and Tanner graph is free of 6-cycles as well.
\end{prof} 

We conclude that if the goal is constructing $3\times n$ exponent matrices with girth 10 whose first column and row are all-zero then by obtaining a difference matrix, $DD$, with disjoint non-zero elements there is no need to consider 6-cycles. And, none of $2\times2$ submatrices are investigated for 4-cycles. Moreover, we generalize the corollary for all $m\times n$ exponent matrices whose first row and column are all-zero as follows.
\begin{Corollary}
	Suppose $B$ is an $m\times n$ exponent matrix. If the first row and column of $B$ are all-zero then the non-existence of 8-cycles proves that equations related to  6-cycle considerations in which the first row of exponent matrix is involved is non-zero.
\end{Corollary} 

According to the above corollary if Tanner graph is 8-cycle free then in order to obtain a Tanner graph with girth 10 there is no need to check  $3\times3$ submatrices whose first row is all-zero.
        
There are three exponent matrices in the literature for $(3,n)$-regular QC-LDPC codes, where $n=4,5,6$, whose Tanner graphs have girth 10 and the lifting degrees obtained are 37, 61 and 91, respectively. Applying the above corollary, we compare the number of computations required to consider in two cases. For the first case we use Fossorier's Lemma by which 1686, 4410, 9555 computations are required. For the second case we use the matrix, $DD$, by which the number of computations are 219, 540 and 1142, respectively. As we see, thanks to the matrix, $DD$, the number of computaions is much less than those when we apply Fossorier's Lemma. Moreover, the number of non-isomorphic QC-LDPC codes obtained by our proposed technique is more than the presented QC-LDPC codes in $\cite{2016}$. We accumulate them in the following table. Furthermore, making use of difference matrices we obtain a (3,7) and (3,8) exponent matrices with girth 10 and lifting degrees $N=145$ and 211. The minimum $N$ achieved in the literature are 159 and 219, respectively.
\begin{table}[h]
	\begin{center}
		\begin{tabular}{|c|c|c|c|c|}
			\hline
			$d_c=4, N=37$ & $\begin{array}{ccc}
			1 & 3 & 24 \\
			11& 33& 5  \\
			\end{array}$ & & & \\
			\hline
			$d_c=5, N=61$&  $\begin{array}{cccc}
			1 & 3 & 21 & 55\\
			5&15&44&31
			\end{array}$ &$\begin{array}{cccc}
			1 & 3 & 21 & 55\\
			14&42&50&38
			\end{array}$&$\begin{array}{cccc}
			1 & 4 & 11 & 27\\
			14&56&32&12
			\end{array}$&$\begin{array}{cccc}
			1 & 4 & 28 & 44\\
			14&56&26&6
			\end{array}$\\
			\hline
				$d_c=6,N=91$&  $\begin{array}{ccccc}
				1 & 6 & 28 & 40&48\\
				10&60&7&36&25
				\end{array}$ &$\begin{array}{ccccc}
				1 & 3&7&25&38\\
				17&51&28&61&9
				\end{array}$& & \\
				\hline
				$d_c=7, N=145$&  $\begin{array}{cccccc}
				1 & 6 & 44 & 71&92&122\\
				4&20&108&83&123&76
				\end{array}$ & & & \\
				\hline
				$d_c=8, N=211$&  $\begin{array}{ccccccc}
				1 & 5 & 12 & 39&56&183&192\\
				3&13&92&188&106&74&151
				\end{array}$ & & & \\
				\hline
		\end{tabular}
		\caption{All non-isomprphic $(3,n)$ QC-LDPC codes with girth 10 and $n=4,5,6$ and an obtained exponent matrix with girth 10 for $n=7,8$, minimum lifting degree obtained in the literature are $N=159,219$, repectively .}
	\end{center}
\end{table}
 
In addition to reducing the search space, making use of the matrix, $DD$, provides us with a chance to obtain a lower bound on the lifting degree of QC-LDPC codes with girth 10 which improves the one in the literature. After the following theorem we show that our proposed lower bound is tight.

\begin{Theorem}
The lifting degree of QC-LDPC codes  with girth 10 is at least $2{n\choose 2}{m\choose 2}+1$.
\end{Theorem}
\begin{prof}
Since the Tanner graphs of $(m,n)$-regular QC-LDPC codes are 8-cycle free, from the second condition of Theorem 1, the difference matrix, $DD$, has no repeated elements. So the matrix $DD$ contains ${m\choose 2}n(n-1)$ disjoint elements. As a result, we have $N\geq 2{n\choose 2}{m\choose 2}+1$.
\end{prof}

The minimum lifting degrees (3,$n$)-regular QC-LDPC codes, $n=4,5,6$, which in the literature are obtained search-based acknowledge our proposed lower bound. They are $N=37,61$ and 91, respectivle. So our proposed lower bound is tight.  

Moreover, the non-existence of repeated elements in the difference matrix, $DD$, to have an  8-cycles free Tanner graph contribute to determine whether a structure can give QC-LDPC code with girth 10 or not. In the following example we demonstrate Tanner graph of QC-LDPC code based on the symmetrical construction presented in $\cite{symmetrical}$ has 8-cycles. 
\begin{Example}
	In $\cite{symmetrical}$ it is shown that if $n$ is an even number then by recognizing the left half of an exponent matrix with girth at least 8, denoted by $A$, it is possible to reach the exponent matrix by  assuming $B=[A\ -A]$. Since the first row of $B$ is all-zero. If $b_{ij}$ and $-b_{ij}$ are two elements of $B$ then $(-b_{ij}, b_{ij})$ and $(b_{ij},-b_{ij})$ are two elements occuring in $DD$. Therefore $DD$ has repeated elements and consequently Tanner graph has 8-cycles.
\end{Example} 
\subsection{10-cycles}\label{}
Nine types of submatrices have to be investigated in order to obtain a QC-LDPC code avoiding 10-cycles. At least three rows  and three columns are required to obtain the left side of Equation (3). So the mentioned submatrices are of sizes $3\times3,3\times4,3\times5,4\times3,4\times4,4\times5,5\times3,5\times4$ and $5\times5$. In this case the matrices, $D$ and $DD$, of the exponent matrix, $B$, are also required. In the following we provide our principle tool to consider 10-cycles.
\begin{lemma}
If the left side of Equation (3), when 10-cycles are under consideration, contains more than two elements of a column of the exponent matrix then its equivalent expression whose elements belong to the difference matrix includes three elements occur in three rows and three columns of $D$. 
\end{lemma}
\begin{prof}
We prove one of types of submatrices which satisfies the condition of Lemma, others are alike and are omitted. Suppose the left side of  Equation (3) is achieved by a $3\times5$ submatrix. For example it is $b_{i_0j_0}-b_{i_0j_1}+b_{i_1j_1}-b_{i_1j_2}+b_{i_2j_2}-b_{i_2j_3}+b_{i_1j_3}-b_{i_1j_4}+b_{i_2j_4}-b_{i_2j_0}$. By rearranging this expression we conclude that 6 terms of the obtained expression are related to 3 elements of the matrix, $D$, which occur in three rows and three columns of $D$. They are $(b_{i_0j_0}-b_{i_2j_0})-(b_{i_0j_1}-b_{i_1j_1})-(b_{i_1j_2}-b_{i_2j_2})=D_{ij_0}-D_{i'j_1}-D_{i''j_2}$. And the other terms are $b_{i_1j_3}-b_{i_2j_3}$ and $-(b_{i_1j_4}-b_{i_2j_4})$ whose equivalent elements of the matrix $D$, $D_{i''j_3}$ and $D_{i''j_4}$, appear in a $2\times2$ submatrix of the matrix, $D$. 
\end{prof}
The advantage of the above Lemma is that we can obtain an equivalence for Equation (3) whose elements belong to the difference matrix, $D$. And more importantly, the elements which are not in the same row or column are investigated in 6-cycle considerations once. So, we can utilise our computations in 6-cycle considerations for investigating 10-cycles.

In the following theorem we provide the necessary and sufficient conditions for a QC-LDPC code whose Tanner graph is free of 10-cycles. The proof is presented in Appendix B. 
\begin{Theorem}
A QC-LDPC code has no 10-cycles if and only if:
\begin{itemize}
\item by choosing every three rows of the exponent matrix, the set consisting of  expressions obtained from the 6-cycle considerations has no intersection by the elements of their three equivalent rows of the matrix, $DD$, and,
\item by choosing every three rows $i,i',i''$ and three columns $j,j',j''$ of the difference matrix, $D$, the set, $X$, consists of expressions $-D_{ij}+D_{i'j'}-D_{i''j''}$ has no common element with a set consisting of the elements of the matrix $DD$, which occur in the rows different from three rows $i,i',i''$. In addition, there is no intersection between the set $X$ and  the values obtained by adding or subtracting every two elements on the distributed diagonals of $2\times2$ submatrices of $D$. Since there are 5 elements for every expression in 10-cycle considerations, the mentioned elements on the distributed diagonal are different from those in the $3\times3$ submatrix of $D$ and,  
\item each $5\times5$ submatrix of the difference matrix, $D$, satisfies the following inequality
\begin{center}
$D_{ij_0}-D_{i'j_1}-D_{i''j_2}-D_{i'''j_3}-D_{i''''j_4}\neq0,$
\end{center}
where $i\neq i'\neq i''\neq i'''\neq i''''\in\{0,1,\dots,{m\choose 2}\}$ and $j_0\neq j_1\neq j_2\neq j_3\neq j_4\in\{0,1,\dots,n-1\}$.
\end{itemize}
\end{Theorem}

\begin{Example}
Suppose
$B=\left[\begin{array}{cccc}
0 & 0 & 0 & 0 \\
0 & 1 & 3 & 13 \\
0 & 9 & 27 & 44 \\
\end{array}\right]$ is the exponent matrix for a $(3,4)$-regular QC-LDPC code with the minimum lifting degree $N=73$ and girth 12. 

The difference matrix is $D=\left[\begin{array}{cccc}
 0 & -1 & -3 & -13 \\
 0 & -9 & -27 & -44 \\
 0 & -8 & -24 & -31 \\
\end{array}\right]$.
And the $3\times6$ difference matrix, $DD$ is presented as follows:
\begin{center}
$\left[\begin{array}{cccccc}
 (1,72) & (3,70) & (13,60) & (2,71) & (12,61) & (10,63)\\
 (9,64) & (27,46) & (29,44) & (18,55) & (35,38) & (17,56) \\
 (8,65) & (24,49) & (31,42) & (16,57) & (23,50) & (7,66) \\
\end{array}\right]$.
\end{center}
The set $A$ consisting of the obtained values from 6-cycle considerations is as follows:
\begin{center}
$A=\{4\ 5\ 11\ 15\ 21\ 22\ 25\ 28\ 30\ 32\ 34\ 37\ 40\ 47\ 53\ 54\ 59\ 67\}$.
\end{center}
And the elements occurring in the matrix, $DD$, is 

$B=\{1\ 2\     3\     7\     8\     9\    10\    12\    13\    16\    17\    18\    23\ 24\    27\    29\    31\    35\    38\    42\    44\    46\    49\    50\    55\    56\ 57\    60\    61\\    63\    64\    65\    66\    70\    71\    72\}$.

Since $A\cap B=\emptyset$ the first condition of theorem above holds for the given exponent matrix. So instead of computing all of equations related to 10-cycle considerations for $3\times3$ and $3\times4$ submatrices, which are 648 equations we consider the non-existence of repeated values in two sets above. It is clear that this manner reduces the complexity of the search algorithm in a great extent. For this exponent matrix just $3\times3$ and $3\times4$ submatrices have to be investigated. As a result, the Tanner graph corresponding to the exponent matrix, $B$, is 10-cycle free.  
\end{Example}
In Appendix B we compute the number of equations required to consider 10-cycles when we use Fossorier's equation and in the following table we summarize them. Note that the number of computations for each type of submatrices is eleven times as many as the number of equations. As we see, the first three submatrices are according to the first condition of theorem above, the second three submatrices are according to the second condition of the theorem  and there is no need to consider the submatrices of sizes $5\times3$ and $5\times4$ when the matrices $D$ and $DD$ are under consideration.
\begin{table}[h]
\begin{center}
\begin{tabular}{|c|c|c|}
\hline
    & Fossorier's equation  & the matrix $DD$ \\
\hline
$3\times3$ submatrices & $54(_{3}^{m})(_{3}^{n})$ & according to the first condition of theorem 5 \\
\hline
$3\times4$ submatrices & $216(_{3}^{m})(_{4}^{n})$ & according to the first condition of theorem 5 \\
\hline
$3\times5$ submatrices & $180(_{3}^{m})(_{5}^{n})$ & according to the first condition of theorem 5 \\
\hline
$4\times3$ submatrices & $72(_{4}^{m})(_{3}^{n})$ & according to the second condition of theorem 5\\
\hline
$4\times4$ submatrices & $864(_{4}^{m})(_{4}^{n})$ & according to the second condition of theorem 5\\
\hline
$4\times5$ submatrices & $960(_{4}^{m})(_{5}^{n})$ & according to the second condition of theorem 5\\
\hline
$5\times3$ submatrices & $540(_{5}^{m})(_{3}^{n})$ &  -\\
\hline
$5\times4$ submatrices & $720(_{5}^{m})(_{4}^{n})$ &  -\\
\hline
$5\times5$ submatrices & $10(_{5}^{m})(_{5}^{n})$  & $10(_{4}^{m})(_{4}^{n})$ \\
\hline
\end{tabular}
\caption{The number of equations  required to check 10-cycles when we use Fossorier's equation and when we utilize the matrix, $DD$}
\end{center}
\end{table}
\begin{Theorem}
If $A$ is a set including all values obtained by 6-cycle considerations and $B$ contains all of values of the matrix, $DD$ then for a regular QC-LDPC code whose Tanner graph is 10-cycle free we have $A\cap B=\emptyset$.
\end{Theorem}
\begin{prof}
As shown in the theorem above if $i,i',i''$ are three rows corresponding to three  rows in 6-cycle considerations, all of values obtained from 6-cycle considerations have no intersection with elements of three rows $i,i',i''$ of the matrix, $DD$. Moreover, the second condition of the theorem shows these values have also no intersection with other rows of the matrix, $DD$. As a result, there are no common elements between two sets.  
\end{prof}
\begin{Corollary}
The minimum lifting degree of a regular QC-LDPC code whose Tanner graph is 10-cycle free is more that $|A|+{m\choose 2}n(n-1)+1$, where $A$ is a set of all values from the left side of equations of Fossorier's Lemma in 6-cycle considerations.
\end{Corollary}
 For example, in the example above $|A|=18$ and ${m\choose 2}n(n-1)+1=37$ so $N\geq55$. From the corollary we conclude that to obtain a regular QC-LDPC code whose Tanner graph is 10-cycle free the minimum lifting degree can be guessed after considering 6-cycles.
\section{Multiple-edge protograph-based QC-LDPC codes }\label{}
In $\cite{Multiple}$, $\cite{2006}$ and  $\cite{voltage}$  it is proved that  $B_{ptg}$ including $i$-component, where $i\geq3$ results in the parity-check matrix whose Tanner Graph has 6-cycles. Also the existence of submatrices like $[2\ 2]$ and $\left[\begin{array}{cc}
2\ 1\\
1\ 1
\end{array}\right]$ of $B_{ptg}$ guarantees the existence of 8-cycles and 10-cycles, respectively.  These cycles are known as inevitable cycles, which we define them in the following. All of the submatrices for inevitable cycles of sizes up to 20 have been investigated in the  literature. 

\begin{Definition}\label{Def3}
	An inevitable cycle of length $2k^{\prime}$ in the Tanner graph of a QC-LDPC code with the lifting degree $N$ is a cycle that always appears in the Tanner graph of the code regardless of how high the lifting degree $N$ is. That is the equation (4) are always equal to zero, regardless of the value of $N$. 
\end{Definition}
\begin{Example}
	Let $m=3$, $n=4$ and $N=13$. Consider the  following exponent matrix for a $(3,4)$-regular multiple-edge QC-LDPC code:
	\begingroup\fontsize{9.5pt}{11pt} \begin{align*}
	B=\left[\begin{array}{cccc}
	(0,1,8) & (\infty) & (0) & (\infty)\\
	(\infty) & (8,12) & (0,4) & (\infty)\\
	(\infty) & (5) & (\infty) & (4,9,10)
	\end{array}\right].
	\end{align*}
	\endgroup
 The following equations I and II  indicate the existence of inevitable cycles of lengths $4$ and $6$ in Tanner graph of the code, respectively.
	\begingroup\fontsize{10pt}{9pt}
	\begin{itemize}
		\item[I.] $B^1_{11}-B^1_{12}+B^2_{12}-B^2_{11}=(8-0)+(4-12)=0$ 
		\item[II.] $B^1_{00}-B^2_{00}+B^3_{00}-B^1_{00}+B^2_{00}-B^3_{00}=(0-1)+(8-0)+(1-8)=0$
	\end{itemize}\endgroup
	 In the following subsection (4-cycles) we will demonstrate that the number of equalities required to be computed in multiple-edge QC-LDPC codes are much more than the ones in single-edge QC-LDPC codes.  Thanks to difference matrices we will define in the following we obtain  the analitical lower bound on the lifting degree of multiple-edge QC-LDPC codes with girth 6. We also construct some exponent matrices to show that the obtained lower bound is tight. Note that, the difference matrices can be used to all $2k$-considerations.
	\begin{Definition}
		Let $B$ be an exponent matrix. For each element of $B$ like $\vec{B}_{ij}$ we take a vector of size $2{|\vec{B}_{ij}|\choose 2}$ as an element of the difference matrix $D$ whose elements are obtained by subtracting every two elemnt of $\vec{B}_{ij}$. If $ij$-th element of $B$ is $(\infty)$ then its equivalent element in $D$ is $(\infty)$ too.  
	\end{Definition}
	\begin{Example}
		Let $B$ be the exponent matrix of Example 5 for which the lifting degree is $N=13$. Its equivalent difference matrix, $D$, is as follows. All of the computations are in modular $N$.
		\begingroup\fontsize{9.5pt}{11pt} \begin{align*}
		D=\left[\begin{array}{cccc}
		(1,12,5,8,6,7) & (\infty) & (0) & (\infty)\\
		(\infty) & (9,4) & (9,4) & (\infty)\\
		(\infty) & (5) & (\infty) & (8,5,7,6,12,1)
		\end{array}\right],
		\end{align*}
		\endgroup	
	\end{Example}
	
	\begin{Definition}
		Let $B$ be an exponent matrix. If $\vec{B}_{ij}$ and $\vec{B}_{i'j}$ are two elements of $j$-the column of $B$ then a vector  $(\vec{B}_{ij},\vec{B}_{i'j})$ whose elements are obtained by subtracting one entry of $\vec{B}_{ij}$ and one entry of $\vec{B}_{i'j}$ is taken as an element of the difference matrix $DD$. It is clear that the vector $(\vec{B}_{ij},\vec{B}_{i'j})$ is of size $|\vec{B}_{ij}|\times|\vec{B}_{i'j}|$. So there is a row for each two rows of $B$. It means that for the exponent matrix of size $m\times n$ the matrix $DD$ is of size ${m\choose2}\times n$. If $ij$-th element of $B$ is $\infty$ and $i'j$-th element is $(\infty)$ or $\vec{B}_{i'j}$ then the vector $(\vec{B}_{ij},\vec{B}_{i'j})$ is $(\infty)$ too. 
	\end{Definition}
	\begin{Example}
		Let $B$ be the exponent matrix of Example 5 for which the lifting degree is $N=13$. Its equivalent difference matrix, $DD$, is as follows.
		\begingroup\fontsize{9.5pt}{11pt} \begin{align*}
		DD=\left[\begin{array}{cccc}
		(\infty) & (\infty) & (0,9) & (\infty)\\
		(\infty) & (\infty) & (\infty) & (\infty)\\
		(\infty) & (3,7) & (\infty) & (\infty)
		\end{array}\right],
		\end{align*}
		\endgroup	
	\end{Example}
\end{Example}
\subsection{4-Cycles}
There are four states to consider all 4-cycles in a multiple-edge QC-LDPC code. They are as follows.
\begin{itemize}
\item[I.] Let $B^z_{ij},B^l_{ij}\in\vec{B}_{ij}$. All submatrices such as $[B^z_{ij},B^l_{ij}]$ have two 1-components in each row and column, when substituting by their corresponding CPMs. So in order to enumerate 4-cycles we can take into consideration such a submatrix as a $2\times2$ submatrix like $\left[\begin{array}{cccc}
B^z_{ij}\ B^l_{ij}\\
B^l_{ij}\ B^z_{ij}
\end{array}\right]$. As a result, if $2(B^z_{ij}-B^l_{ij})=0\ (\mod\ N)$ then there are 4-cycles.
\item[II.] Let $B^z_{ij},B^l_{ij}\in\vec{B}_{ij}$ and $B^{z'}_{ik},B^{l'}_{ik}\in\vec{B}_{ik}$. If we have $[(B^z_{ij}\ B^l_{ij})\ (B^{z'}_{ik}\ B^{l'}_{ik})]$ as a submatrix, then the equality $\pm(B^z_{ij} - B^l_{ij})\pm(B^{z'}_{ik} - B^{l'}_{ik})=0\ (\mod\ N)$  results in 4-cycles. 
\item[III.] Let $B^z_{ij},B^l_{ij}\in\vec{B}_{ij}$ and $B^{z'}_{kj},B^{l'}_{kj}\in\vec{B}_{kj}$. We have to apply Equation (4) for all submatrices such as $\left[\begin{array}{cc}
(B^z_{ij}\ B^l_{ij})\\
(B^{z'}_{kj}\ B^{l'}_{kj})
\end{array}\right]$. If $\pm (B^z_{ij}-B^l_{ij})\pm( B^{z'}_{kj}-B^{l'}_{kj})=0\ (\mod\ N)$ then the Tanner graph has 4-cycles.
\item[IV.] We have to apply Equation (4) for all submatrices such as $\left[\begin{array}{cc}
\vec{B}_{ij}\ \vec{B}_{ik}\\
\vec{B}_{i'j}\ \vec{B}_{i'k}
\end{array}\right]$. This submatrix includes ${(|\vec{B}_{ij}|\times|\vec{B}_{ik}|\times|\vec{B}_{i'j}|\times|\vec{B}_{i'k}|)}$, $2\times2$ submatrices for which we have to consider 4-cycles.
\end{itemize}

In the gollowing, taking benefits of two difference matrices $D$ and $DD$ we investigate the necessary and sufficient conditions for a multiple-edge QC-LDPC code to have girth at least 6.
\begin{Theorem}
	A multiple-edge QC-LDPC code is free of 4-cycles if and only if
	\begin{itemize}
		\item $2\times D\ (\mod\ N)$ contains no zero element,
		\item There are no repeated elements in each row and each column of $D$,
		\item Each row of $DD$ is free of repeated elements.
	\end{itemize}
\end{Theorem}
\begin{prof}
In the state I in 4-cycle considerations we have $2(B^z_{ij}-B^l_{ij})=0$. Since $B^z_{ij},B^l_{ij}\in\vec{B}_{ij}$ we conclude that $B^z_{ij}-B^l_{ij}$ is an entry of the $ij$-the vector of the difference matrix, $D$, corresponding to $\vec{B}_{ij}$. So in order to construct multiple-edge QC-LDPC codes with girth at least 6 it is necessary to have a difference matrix, $D$, for which the matrix $2D$ is free of zero elements.

From the state II we conclude that if girth is at least 6 then each row of the difference matrix, $D$, is free of repeated elements. Because, $\pm(B^z_{ij} - B^l_{ij})$ and $\pm(B^z_{ik} - B^l_{ik})$ occur in the vectors of $D$ which are corresponding to the vectors $\vec{B}_{ij}$ and $\vec{B}_{ik}$ of $B$, respectively. similarly, from the state III we conclude that if girth is at least 6 then each column of the difference matrix, $D$, is free of repeated elements.

As mentioned in the state IV, for a submatrix of $B$ such as $\left[\begin{array}{cc}
\vec{B}_{ij}\ \vec{B}_{ik}\\
\vec{B}_{i'j}\ \vec{B}_{i'k}
\end{array}\right]$ 
there are\\ $(|\vec{B}_{ij}|+|\vec{B}_{ik}|+|\vec{B}_{i'j}|+|\vec{B}_{i'k}|)$, $2\times2$ submatrices to investigate. Suppose $B^l_{ij}\in\vec{B}_{ij}, B^{l'}_{ik}\in\vec{B}_{ik}, B^z_{i'j}\in\vec{B}_{i'j}$ and $B^{z'}_{i'k}\in\vec{B}_{i'k}$ are elements of a $2\times2$ submatrix. Then the equality $B^l_{ij}-B^{l'}_{ik}+B^{z'}_{i'k}-B^z_{i'j}=0\ \ (\mod\ N)$ or equivalently the equality $B^l_{ij}-B^z_{i'j}=B^{l'}_{ik}-B^{z'}_{i'k}\ \ (\mod\ N)$ gives 4-cycles. Since $B^l_{ij}-B^z_{i'j}$ and $B^{l'}_{ik}-B^{z'}_{i'k}$  occur in a row of the difference matrix, $DD$, we conclude that repeated elements in a row of $DD$ gives rise to 4-cycles. 
\end{prof}

One of the most important advantages of taking two difference matrices $D$ and $DD$ is the lower bound of the lifting degree which is analyticaly obtained. 
\begin{Theorem}
	Let $B$ be the exponent matrix of a multiple-edge QC-LDPC code of size $m\times n$ with girth at least 6. We take three variables as follows.
	
	\noindent  $A=\max\{2\sum_{j=1}^{n}{|\vec{B}_{ij}|\choose2};i=1,2\dots,m\}$, $B=\max\{2\sum_{i=1}^{m}{|\vec{B}_{ij}|\choose2};j=1,2\dots,n\}$ and
	
	\noindent $C=\max\{\sum_{j=1}^{n}|\vec{B}_{ij}|\times|\vec{B}_{i'j}|;i\neq i';\ i,i'\in\{1,2\dots,m\}\}$. 
	
	The minimum lifting degree is at least $N=\max\{A,B,C\}$.
\end{Theorem}
\begin{prof}
	According to the previous theorem, since girth is at least 6  every row of the difference matrix, $D$, has no repeated elements. So $A$ is the maximum number of elements occuring in a row of $D$. Similarly, $B$ is the maximum number of elements appearing in a column of $D$. In addition, girth is at least 6 so in the difference matrix, $DD$, every row is without repaeted elemnts. Therefore, the maximum number of elements in a row of $DD$ is $C$. As a result, the minimum lifting degree is at least the maximum of $A$, $B$ and $C$. 
\end{prof}
\begin{Example}
	Let $B$ be the exponent matrix of a multiple-edge $(4,2n)$-regular QC-LDPC code with girth at least 6 whose elements in the corresponding base matrix is 2. Then every element in the difference matrix, $D$ is a vector of size 2. Therefore, there are $2n$ elements in each row of the difference matrix $D$. As a result $A=2n$. Also the number of elemnts in each column of $D$ is 4, so $|B|=4$. The difference matrix, $DD$, is a row with $n$ vectors of size 4, hence, $|C|=4n$.   Consequently, $N\geq 4n$.
	\end{Example}
	In this paper, we present  exponent matrices of  multiple-edge $(4,2n)$-regular QC-LDPC codes with girth 6 and $N=4n$, where $2\leq n\leq 6$. These matrices demonstrate that the obtained lower bound of the lifting degree in  Theorem 6 is tight. In the following table we compare the minimum lifting degrees and also the lengths of both single-edge and multiple-edge QC-LDPC codes with the same row and column weight. It shows that the length of multiple-edge $(4,2n)$-regular QC-LDPC codes with girth 6 is much less than their counterparts. 
\begin{table}[h]
	\begin{center}
		\begin{tabular}{|c|c|c|c|}
			\hline
			column weight & exponent matrix of multiple-edge QC-LDPC code & multiple-edge & single-edge\\ 
			\hline
				$d_c=4$ &  $B=\left[\begin{array}{ccc}
				(0\ 1) & (0\ 2)\\
				(1\ 3) & (6\ 7)
				\end{array}\right]$ & $N=8$, length=16 & -\\
				\hline
				 $d_c=6$ &  $B=\left[\begin{array}{ccccc}
				 (0\ 1) & (0\ 2) & (0\ 4) \\
				 (1\ 5) & (8\ 11) & (2\ 7)
				 \end{array}\right]$ & $N=12$, length=36 & $N=7$, length=42\\
				 \hline
				  $d_c=8$ &  $B=\left[\begin{array}{ccccc}
				  (0\ 1) & (0\ 2) & (0\ 3) & (0\ 4)\\
				  (1\ 3) & (6\ 15) & (8\ 12) & (11\ 14)
				  \end{array}\right]$ & $N=16$, length=64 & $N=10$, length=80\\
				 \hline
				   $d_c=10$ &  $B=\left[\begin{array}{ccccc}
				   (0\ 1) & (0\ 2) & (0\ 3) & (0\ 4) & (0\ 5)\\
				   (1\ 3) & (6\ 9) & (11\ 16) & (18\ 19) & (10\ 17)
				   \end{array}\right]$ & $N=20$, length=100 & $N=11$, length=110\\
				  \hline
				    $d_c=12$ &  $B=\left[\begin{array}{cccccc}
				    (0\ 1) & (0\ 2) & (0\ 3) & (0\ 4) & (0\ 5) & (0\ 6) \\
				    (1\ 3) & (6\ 7) & (13\ 21) & (12\ 23) & (14\ 20) & (17\ 22)
				    \end{array}\right]$ & $N=24$, length=144 & $N=13$, length=156\\
				   \hline
		\end{tabular}
		\caption{Comparison of length of Single-edge and Multiple-edge QC-LDPC codes with the same column and row weights.}
	\end{center}
\end{table}
\subsection{6-cycles}
There are seven states to consider all 6-cycles in a multiple-edge QC-LDPC code. We, first, explain them in the following. And in the next,  taking benefit of two difference matrices $D$ and $DD$ we present the necessary and sufficient conditions of an exponent matrix to be 6-cycle free. 
\begin{itemize}
	\item[I.] Let $B^z_{ij},B^l_{ij}\in\vec{B}_{ij}$. If $3(B^z_{ij}-B^l_{ij})=0\ (\mod\ N)$ then there are 6-cycles.
	\item[II.] Let $B^z_{ij},B^l_{ij}\in\vec{B}_{ij}$ and $B^{z'}_{ij'},B^{l'}_{ij'}\in\vec{B}_{ij'}$. If we have $[(B^z_{ij}\ B^l_{ij})\ (B^{z'}_{ik}\ B^{l'}_{ik})]$ as a submatrix, then $\pm2(B^z_{ij}-B^l_{ij})\pm(B^{z'}_{ij'}-B^{l'}_{ij'})=0\ (\mod\ N)$ or $\pm(B^z_{ij}-B^l_{ij})\pm2(B^{z'}_{ij'}-B^{l'}_{ij'})=0\ (\mod\ N)$  results in 6-cycles. 
	\item[III.] Let $B^z_{ij},B^l_{ij}\in\vec{B}_{ij}$ and $B^{z'}_{i'j},B^{l'}_{i'j}\in\vec{B}_{i'j}$. We have to apply equation (4) for all submatrices such as $\left[\begin{array}{cc}
	(B^z_{ij}\ B^l_{ij})\\
	(B^{z'}_{i'j}\ B^{l'}_{i'j})
	\end{array}\right]$. If $\pm 2(B^z_{ij}-B^l_{ij})\pm( B^{z'}_{i'j}-B^{l'}_{i'j})=0\ (\mod\ N)$ or $\pm (B^z_{ij}-B^l_{ij})\pm2( B^{z'}_{i'j}-B^{l'}_{i'j})=0\ (\mod\ N)$ then the Tanner graph has 6-cycles.
	\item[IV.] Let $B^z_{ij},B^l_{ij}\in\vec{B}_{ij}$, $B^{z'}_{ij'},B^{l'}_{ij'}\in\vec{B}_{ij'}$,   $B^{z''}_{i'j},B^{l''}_{i'j}\in\vec{B}_{i'j}$ and $B^{z'''}_{i'j'},B^{l'''}_{i'j'}\in\vec{B}_{i'j'}$. If we have $\left[\begin{array}{ccc}
	(B^z_{ij}\ B^l_{ij}) && (B^{z'}_{ij'}\ B^{l'}_{ij'})\\
	(B^{z''}_{i'j}\ B^{l''}_{i'j}) && (B^{z'''}_{i'j'}\ B^{l'''}_{i'j'})
	\end{array}\right]$ as a submatrix then equalities like  $\pm(B^z_{ij}-B^l_{ij})+B^z_{ij}-B^{z'}_{ij'}=B^{z''}_{i'j}-B^{l'''}_{i'j'}\ (\mod\ N)$ gives rise to 6-cycles. The number of such equations, for each submatrix shown above, is 64.
	\item[V.] Let $B^z_{ij},B^l_{ij}\in\vec{B}_{ij}$, $B^{z'}_{ij'},B^{l'}_{ij'}\in\vec{B}_{ij'}$ and  $B^{z''}_{ij''},B^{l''}_{ij''}\in\vec{B}_{ij''}$. $\pm(B^z_{ij}-B^l_{ij})\pm(B^{z'}_{ij'}-B^{l'}_{ij'})\pm(B^{z''}_{ij''}-B^{l''}_{ij''})=0\ (mod\ N)$  results in 6-cycles if we have $[(B^z_{ij}\ B^l_{ij})\ (B^{z'}_{ij'}\ B^{l'}_{ij'})\ (B^{z''}_{ij''}\ B^{l''}_{ij''})]$ as a submatrix. 
	\item[VI.] Let $B^z_{ij},B^l_{ij}\in\vec{B}_{ij}$, $B^{z'}_{ij'},B^{l'}_{ij'}\in\vec{B}_{ij'}$, $B^{z''}_{ij''},B^{l''}_{ij''}\in\vec{B}_{ij''}$,   $B^{z'''}_{i'j},B^{l'''}_{i'j}\in\vec{B}_{i'j}$, $B^{z^4}_{i'j'},B^{l^4}_{i'j'}\in\vec{B}_{i'j'}$ and $B^{z^5}_{i'j''},B^{l^5}_{i'j''}\in\vec{B}_{i'j''}$. If we have $\left[\begin{array}{ccccc}
	(B^z_{ij}\ B^l_{ij}) && (B^{z'}_{ij'}\ B^{l'}_{ij'}) && (B^{z''}_{ij''}\ B^{l''}_{ij''})\\
	(B^{z'''}_{i'j}\ B^{l'''}_{i'j}) && (B^{z^4}_{i'j'}\ B^{l^4}_{i'j'}) && (B^{z^5}_{i'j''}\ B^{l^5}_{i'j''})
	\end{array}\right]$ as a submatrix then equalities like  $\pm(B^z_{ij}-B^l_{ij})+B^{z'}_{ij'}-B^{l''}_{ij''}=B^{l^4}_{i'j'}-B^{z^5}_{i'j''}$ give rise to 6-cycles. The number of such equations, for each submatrix shown above, is 96.
	\item[VII.] We have to apply equation (4) for all submatrices such as $\left[\begin{array}{ccccc}
	\vec{B}_{ij}  && \vec{B}_{ij'} && \vec{B}_{ij''}\ \\
	\vec{B}_{i'j} && \vec{B}_{i'j'} && \vec{B}_{i'j''}\\
	\vec{B}_{i''j} && \vec{B}_{i''j'} && \vec{B}_{i''j''}\\
	\end{array}\right]$. The number of $3\times3$ submatrices in this submatrix is the product of cardinality of all entries in this submatrix.
\end{itemize}
In the following theorem we consider 6-cycles making use of difference matrices.
\begin{Theorem}
	A multiple-edge QC-LDPC code is free of 6-cycles if and only if 
	\begin{itemize}
		\item $3\times D\ (\mod\ N)$ contains no zero element,
		\item If $i-$th row of the difference matrix $D$ and $2D$ are shown by $D_i$ and $2D_i$, respectivly, then $D_i\cap 2D_i=\emptyset$. Similarly, for $j$-th column of these two matrices we have $D_j\cap 2D_j=\emptyset$.
		\item If $d_{ij}\in\vec{D}_{ij},\ d_{ij'}\in\vec{D}_{ij'}$ and $d_{ij''}\in\vec{D}_{ij''}$ are elements of the difference matrix, $D$, then $\pm d_{ij}\pm d_{ij'}\pm d_{ij''}\neq0\ (\mod\ N)$.
		\item Suppose two rows $i$ and $i'$ and two columns $j$ and $j'$ of the exponent matrix are under consideration. If their corresponding vectors in the difference matrix, $DD$, are $\vec{DD}_{(ii')j}$, $\vec{DD}_{(ii')j'}$ then construct a set $A$, with cardinality  $|\vec{DD}_{(ii')j}||\vec{DD}_{(ii')j'}|$, whose elements are obtained by subtracting an element of $DD_{(ii')j}$ and an element of $DD_{(ii')j'}$. If Tanner graph has no 6-cycle then $A\cap D_i=\emptyset$ and $A\cap D_{i'}=\emptyset$.
		\item For each $3\times3$ submatrix of the exponent matrix Equation (4) is nonzero.
	\end{itemize}
\end{Theorem}
\begin{prof}
	From the state $I$ we have the first item. The states $II$ annd $III$ give the second item. The state $V$ is equavalent to the third item. States $IV$ and $VI$ are simlified in the forth item. Finaly the state $VII$ and the fifth item are equivalent.
\end{prof}

We apply Theorem 7 to obtain  multiple-edge $(4,2n)$-regular QC-LDPC code with girth 8 and $n=2,3,4$ presented in the following table. The minimum lifting degree obtained is $N=30,50$ and 78 respectively.
\begin{table}[h]
	\begin{center}
		\begin{tabular}{|c|c|c|c|}
			\hline
			$d_c=4$ & $d_c=6$ & $d_c=8$ \\
			\hline
			  $B=\left[\begin{array}{cc}
			(0\ 1) & (0\ 22)\\
			(0\ 11) & (15\ 17)
			\end{array}\right]$ &  $B=\left[\begin{array}{ccc}
			(0\ 1) & (0\ 47) & (0\ 8) \\
			(0\ 5) & (15\ 36) & (30\ 32)
			\end{array}\right]$ &   $B=\left[\begin{array}{cccc}
			(0\ 1) & (0\ 27) & (0\ 37) & (0\ 33)\\
			(0\ 4) & (21\ 61) & (29\ 52) & (18\ 46)
			\end{array}\right]$\\
			\hline
		\end{tabular}
		\caption{ Multiple-edge $(4,2n)$ QC-LDPC codes with girth 8 and $n=2,3,4$.}
	\end{center}
\end{table}

\section{Conclusion}\label{}
Many efforts have been put into constructing quasi-cyclic low density parity-check (QC-LDPC) codes with large girths, shortest lengths for a specific degree distribution. The property of the shortest length has been investigated for QC-LDPC codes lifted from fully-connected base graphs known as single-edge protographs. Moreover, there are some lower bounds on the lifting degree of QC-LDPC codes with different girths. In this paper, we take an $m\times n$ exponent matrix, $B$, then define two matrices named as ``difference matrices", denoted by $D$ and  $DD$ which  contribute to reduce the complexity of search algorithms to achieve a regular QC-LDPC code with the shortest length and the certain girth. More importantly, making use of these matrices we demonstrate that the necessary and sufficient condition for the difference matrix $D$ of a QC-LDPC code to have a Tanner graph with girth 6 is that every row is free of repeated elements.  In addition, we prove that if the elements of the first row and the first column of the exponent matrix are zeros and Tanner graph is free of 8-cycles then in order to achieve a Tanner graph with girth 10  there is no need to check 6-cycle equalities on $3\times3$ submatrices containing the first row.  Furthermore, making use of these matrices we provide some lower bounds on the lifting degree, denoted by $N$, of a regular QC-LDPC code with specific column and row weights and the desired girth. In fact, we prove that if an exponent matrix is of size $m\times n$ then for QC-LDPC codes whose Tanner graphs have the girth of 10 and 12 we have $N\geq {m\choose 2}n(n-1)+1$ and $N\geq |A|+{m\choose 2}n(n-1)+1$, where $A$ is a set of all values from the left side of equations of Fossorier's Lemma in 6-cycle considerations, respectively. The lower bound when girth is 10 is improved in comparison with the one in the literature. Applying difference matrices we obtain the exact number of non-isomorphic $(3,n)$ QC-LDPC codes with girth 10 for $n=5,6$. Also for $n=7,8$ we obtain a lifting degree $N=145$ and 211, respectively, which are less than those provided in the literature, $N=159$ and 219, respectively. 

We also attempt to achieve some usefull results related to both regular and irregular multiple-edge QC-LDPC codes with girth 6 and 8. We demonstrate a tight lower bound on the lifting degree when girth is 6 along with some exponent matrices to show the tightness of our proposed bound. In fact, we  analytically prove that if $\vec{B}_{ij}$ is $ij$-th element of the exponent matrix $B$ then by taking three values  $A=\max\{2\sum_{j=1}^{n}{|\vec{B}_{ij}|\choose2};i=1,2\dots,m\}$, $B=\max\{2\sum_{i=1}^{m}{|\vec{B}_{ij}|\choose2};j=1,2\dots,n\}$ and $C=\max\{\sum_{j=1}^{n}|\vec{B}_{ij}|\times|\vec{B}_{i'j}|;i\neq i'; i,i'\in\{1,2\dots,m\}\}$  the minimum lifting degree is at least $N=\max\{A,B,C\}$. Another consequence is regarding to reduction in the equalities required to be investigated when Tanner graph with girth 8 is desired. For this case we present some exponent matrices too.

%%%%%%%%%%%%%%%%%%%%%%%%%%%%%%%%%%%%%%%%%%%%%%

{\bf APPENDIX A}

In this appendix we simultaneously  prove the theorem 4 and compute the number of equations for each submatrix when 8-cycles are under consideration. In order to prove we divide 8-cycle investigations into three cases. In the first case all submatrices contain two rows of the exponent matrix. In the second case they include three rows and for the third case there are four rows for each submatrix. Moreover, for each case there are three items to consider.
\begin{prof}
{\bf Case 1}: Submatrices which include two rows of the exponent matrix. Suppose they are $i_0$ and $i_1$.
\begin{enumerate} 
\item  If two columns are $j_0,j_1$ then the left side of Equation (1) is 

$b_{i_0j_0}-b_{i_0j_1}+b_{i_1j_1}-b_{i_1j_0}+b_{i_0j_0}-b_{i_0j_1}+b_{i_1j_1}-b_{i_1j_0}=2(b_{i_0j_0}-b_{i_0j_1}+b_{i_1j_1}-b_{i_1j_0})$.

By rearranging the equation we obtain the following relation. $$2((b_{i_0j_0}-b_{i_1j_0})-(b_{i_0j_1}-{i_1j_1}))=2(D_{ij_0}-D_{ij_1}),$$ where $i\in\{0,1,\dots, {m\choose 2}\}$. So, in the difference matrix of 8-cycle free QC-LDPC codes we have $2(D_{ij_0}-D_{ij_1})\neq0$. Since $(D_{ij_0}-D_{ij_1})$ is a component of an element of the matrix, $DD$, in 8-cycle free QC-LDPC codes the matrix, $2DD$, has no zero element. So the first condition of the theorem is proved by this item.

In this item if Fossorier's equation is used then ${m\choose 2}{n\choose 2}$ equations are required to investigate. And for each one there are 9 operations to compute. So the number of computations is $9{m\choose 2}{n\choose 2}$. Whereas, by the matrix $2DD$ the number of operations is ${m\choose 2}{n\choose 2}$. 
\item  Take $j_0,j_1,j_2$ as three columns of submatrices. The left side of Equation (1) and its corresponding equation whose elements belong to $D$ are as follows:

$b_{i_0j_0}-b_{i_0j_1}+b_{i_1j_1}-b_{i_1j_2}+b_{i_0j_2}-b_{i_0j_1}+b_{i_1j_1}-b_{i_1j_0}=(b_{i_0j_0}-b_{i_1j_0})-(b_{i_0j_1}-b_{i_1j_1})+(b_{i_0j_2}-b_{i_1j_2})-(b_{i_0j_1}-b_{i_1j_1})=D_{ij_0}-D_{ij_1}+D_{ij_2}-D_{ij_1}$.

So, in the difference matrix of an 8-cycle free QC-LDPC code we have $D_{ij_0}-D_{ij_1}+D_{ij_2}-D_{ij_1}\neq0$. Or equivalently, $D_{ij_0}-D_{ij_1}\neq D_{ij_1}-D_{ij_2}$. To obtain this inequality the 8-cycle is started from $b_{i_0j_0}$. If the cycle is started from $b_{i_0j_1}$ then we obtain one of the inequalities $D_{ij_0}-D_{ij_1}\neq -(D_{ij_0}-D_{ij_2})$ or $D_{ij_0}-D_{ij_2}\neq -(D_{ij_1}-D_{ij_2})$. Note that the two sides of inequalities are components of two elements of $i$-th row of the matrix $DD$. 

Therefore, in this item if Fossorier's equation is used then $3{m\choose 2}{n\choose 3}$ equations  are required to investigate. And  $27{m\choose 2}{n\choose 3}$ operations are computed.
\item  Consider four columns $j_0,j_1,j_2$ and $j_3$. Like the previous item by investigating Equations (1) and their equivalences in the difference matrix of 8-cycle free QC-LDPC codes we have: $\pm(D_{ij_0}-D_{ij_1})\neq \pm(D_{ij_2}-D_{ij_3})$. Note that the two sides of inequalities are components of two elements of $i$-th row of the matrix $DD$. 

In this item if Fossorier's equation is used then $6{m\choose 2}{n\choose 4}$ equations are required to investigate.  The number of computations is $54{m\choose 2}{n\choose 4}$.

The items 2 and 3 result in the existence of $2{m\choose 2}$ disjoint elements for each row of $DD$. Suppose $(D_{ij}-D_{ij'},D_{ij'}-D_{ij})$ and $(D_{ij}-D_{ij''},D_{ij''}-D_{ij})$ are two elements of $i$-th row of $DD$. If they have elements in common then one of the following equalities occurs which contradicts the assumption that the Tanner graph is 8-cycle free. The equalities are $D_{ij'}=D_{ij''}$ or $(D_{ij}-D_{ij'})+D_{ij}-D_{ij''}=0$. The item 3  proves that the equality $\pm(D_{ij}-D_{ij'})=\pm(D_{ij''}-D_{ij'''})$ provides 8-cycles. 

 {\bf Case 2}: Submatrices which include three rows of the exponent matrix. Suppose they are $i_0,\ i_1$ and $i_2$.

\item  Suppose two columns are $j_0,j_1$. The left side of Equations (1) and their corresponding equations whose elements belong to $D$ are as follows:

$1)\ b_{i_0j_0}-b_{i_0j_1}+b_{i_1j_1}-b_{i_1j_0}+b_{i_2j_0}-b_{i_2j_1}+b_{i_1j_1}-b_{i_1j_0}=(b_{i_0j_0}-b_{i_1j_0})-(b_{i_0j_1}-b_{i_1j_1})-(b_{i_1j_0}-b_{i_2j_0})+(b_{i_1j_1}-b_{i_2j_1})=D_{ij_0}-D_{ij_1}-D_{i'j_0}+D_{i'j_1}$,

$2)\ b_{i_1j_0}-b_{i_1j_1}+b_{i_0j_1}-b_{i_0j_0}+b_{i_2j_0}-b_{i_2j_1}+b_{i_0j_1}-b_{i_0j_0}=-(b_{i_0j_0}-b_{i_1j_0})+(b_{i_0j_1}-b_{i_1j_1})-(b_{i_0j_0}-b_{i_2j_0})+(b_{i_0j_1}-b_{i_2j_1})=-D_{ij_0}+D_{ij_1}-D_{i'j_0}+D_{i'j_1}$,

$3)\ b_{i_1j_0}-b_{i_1j_1}+b_{i_2j_1}-b_{i_2j_0}+b_{i_0j_0}-b_{i_0j_1}+b_{i_2j_1}-b_{i_2j_0}=(b_{i_0j_0}-b_{i_2j_0})-(b_{i_0j_1}-b_{i_2j_1})+(b_{i_1j_0}-b_{i_2j_0})-(b_{i_1j_1}-b_{i_2j_1})=D_{ij_0}-D_{ij_1}+D_{i'j_0}-D_{i'j_1}$.

So, by considering all of conditions above, for 8-cycle free QC-LDPC codes we have: $\pm(D_{ij_0}-D_{ij_1})\neq \pm(D_{i'j_0}-D_{i'j_1})$, where $i\neq i'; i,i'\in\{0,1,\dots, {m\choose 2}\}$. Thus, in any column of the matrix $DD$ there are no repeated elements.

In this item if Fossorier's equation is used then $3(_{3}^{m})(_{2}^{n})$ equations are required to investigate. And the number of computations is $27(_{3}^{m})(_{2}^{n})$.
\item Take $3\times3$ submatrices of the exponent matrix, where $j_0,\ j_1$ and $j_2$  are their three columns. Like the previous item by investigating  Equations (1), which contains 9 cases, and their equivalences in the difference matrix we have: $\pm(D_{ij_0}-D_{ij_1})\neq \pm(D_{i'j_1}-D_{i'j_2})$.

In this item if Fossorier's equation is used then $18(_{3}^{m})(_{3}^{n})$ equations are required to investigate. 
\item  Take $3\times4$ submatrices of the exponent matrix, where $j_0,\ j_1,\ j_2$ and $j_3$  are the four columns. The left side of Equation  (1) is $$b_{i_0j_0}-b_{i_0j_1}+b_{i_1j_1}-b_{i_1j_2}+b_{i_2j_2}-b_{i_2j_3}+b_{i_1j_3}-b_{i_1j_0}.$$ In the following we rearrange the expression to obtain the one whose terms are elements of $D$.

$(b_{i_0j_0}-b_{i_1j_0})-(b_{i_0j_1}-b_{i_1j_1})-(b_{i_1j_2}-b_{i_2j_2})+(b_{i_1j_3}-b_{i_2j_3})=(D_{ij_0}-D_{ij_1})-(D_{i'j_2}-D_{i'j_3})$.
In this case elements in the first term occur in $i$-th row of $D$ and elements in the second term occur in $i'$-th row of $D$. As a result, if the  Tanner graph is 8-cycle free then by taking each four columns in two rows of $D$ we have $\pm(D_{ij_0}-D_{ij_1})\neq \pm(D_{i'j_2}-D_{i'j_3})$.

In this item if Fossorier's equation is used then $6{m\choose 3}{n\choose 4}$ equations are required to investigate. The number of computations is $54{m\choose 3}{n\choose 4}$

 {\bf Case 3}: Submatrices which include four rows of the exponent matrix. Suppose they are $i_0,\ i_1,\ i_2$ and $i_3$.

\item Suppose two columns are $j_0,j_1$. The left side of Equations (1) and their corresponding equations whose elements belong to $D$ are similar to:

$b_{i_0j_0}-b_{i_0j_1}+b_{i_1j_1}-b_{i_1j_0}+b_{i_2j_0}-b_{i_2j_1}+b_{i_3j_1}-b_{i_3j_0}=D_{ij_0}-D_{ij_1}+D_{i'j_0}-D_{i'j_1}$.

This case is similar to considering $3\times2$ submatrices. By considering all of them we obtain the same result. So, for 8-cycle free QC-LDPC codes we have: $\pm(D_{ij_0}-D_{ij_1})\neq \pm(D_{i'j_0}-D_{i'j_1})$, where $i\neq i'; i,i'\in\{0,1,\dots, (_{2}^{m})\}$.

In this item if Fossorier's equation is used then $6{m\choose 4}{n\choose 2}$ equations are required to investigate. Whereas, since this item is considered in $3\times2$ submatrices, there is no need to investigate again. 
\item  Take $4\times3$ submatrices of the exponent matrix, where $j_0,\ j_1$ and $j_2$  are their three columns. By investigating all equations we conclude that for 8-cycle free QC-LDPC codes the following inequality holds: $\pm(D_{ij_0}-D_{ij_1})\neq \pm(D_{i'j_1}-D_{i'j_2})$. So, this case is similar to considering $3\times3$ submatrices and is not investigated if the matrix, $D$ is under consideration. However, if Fossorier's equation is used then $12(_{4}^{m})(_{3}^{n})$ equations are required to investigate.

 The items 4 to 8 result in the non-existence of repeated elements for each two rows of $DD$. As a whole, the difference matrix, $DD$, belonging to 8-cycle free QC-LDPC codes has no repeated entries. Consequently, from the items 2 to 8 the second condition of the theorem is proved.
 
\item  If four columns are $j_0,j_1,j_2,j_3$ then Equation (1) is $b_{i_0j_0}-b_{i_0j_1}+b_{i_1j_1}-b_{i_1j_2}+b_{i_2j_2}-b_{i_2j_3}+b_{i_3j_3}-b_{i_3j_0}.$ In the following we rearrange the expression to obtain the one whose terms are compared with elements of  $D$. 

 $(b_{i_0j_0}-b_{i_3j_0})-(b_{i_0j_1}-b_{i_1j_1})-(b_{i_1j_2}-b_{i_2j_2})-(b_{i_2j_3}-b_{i_3j_3})=D_{ij_0}-D_{i'j_1}-D_{i''j_2}-D_{i'''j_3}.$ 
 
So in this case all elements of a $4\times4$ submatrix of $D$ are considered. If for every four elements of the $4\times4$ submatrix, none of which occurs in the same row and the same column, the above expression is non-zero modular $N$ then such submatrices do not provide 8-cycles.
 The number of Fossorier's equation to consider is $8{m\choose 4}{n\choose 4}$. From this item we conclude the third condition of the theorem.
\end{enumerate}

\end{prof}

{\bf APPENDIX B}

In this appendix we simultaneously  prove the theorem 5 and compute the number of equations for each submatrix when 10-cycles are under consideration. In order to prove we divide 10-cycle investigations into three cases. In the first case all submatrices contain three rows of the exponent matrix. In the second case they include four rows and for the third case there are five rows for each submatrix. Moreover, for each case there are three items to consider.
\begin{prof} 
{\bf Case 1}: Submatrices which include three rows of the exponent matrix. Suppose they are $i_0,\ i_1$ and $i_2$. In this case three elements of the difference matrix, which occur in three columns and three rows, appear in expressions related to 6-cycle considerations.
\begin{enumerate}
\item  If three columns are $j_0,j_1,j_2$ then its equivalent $3\times3$ submatrix, $D'$, of the difference matrix, $D$, consists of three rows $i,i',i''$ and three columns $j_0,j_1,j_2$. Elements of the exponent matrix which are investigated for 6-cycles are on a distributed diagonal of $D'$. For each three elements on a distributed diagonal of $D'$ there are nine $2\times2$ submatrices of $B$ which are two elements in one row of $D'$. According to the definition of the difference matrix, $DD$, these two elements are one element of $DD$. So, instead of considering all nine equations of Fossorier's Lemma for each distributed diagonal, which are 54 equations as a whole, we take benefit our results in 6-cycle consideration and the matrix, $DD$. In the following we provide the left side of Equation (1) for one of those 54 cases and its equivalent equation from the matrices, $D$ and $DD$. 

$b_{i_1j_0}-b_{i_1j_2}+b_{i_2j_2}-b_{i_2j_1}+b_{i_0j_1}-b_{i_0j_2}+b_{i_1j_2}-b_{i_1j_1}+b_{i_0j_1}-b_{i_0j_0}=-(b_{i_0j_0}-b_{i_1j_0})+(b_{i_0j_1}-b_{i_2j_1})-(b_{i_1j_2}-b_{i_2j_2})+(b_{i_0j_1}-b_{i_1j_1})-(b_{i_0j_2}+b_{i_1j_2})=-D_{ij_0}+D_{i'j_1}-D_{i''j_2}+D_{ij_1}-D_{ij_2}=-D_{ij_0}+D_{i'j_1}-D_{i''j_2}+DD_{ij}$. Where, $j$ is a column of the matrix, $DD$.
In this item if Fossorier's equation is used then $54{n\choose 3}{m\choose 3}$ equations are required to investigate.

\item  Take $j_0,j_1,j_2,j_3$ as four columns of a $3\times4$ submatrix. For this item there are $6(_{3}^{4})9=216$ cases to consider. Because, there are $(_{3}^{4})$ choices for obtaining $3\times3$ submatrices and for each of them there are 6 considerations related to 6-cycles. The number of mentioned $2\times2$ submatrices are $(_{2}^{3})\times(_{1}^{3})$. The left side of Equation (1) related to one of them is presented in the following.  The right side of the equation consists of the elements of the matrix $D$.

$b_{i_1j_0}-b_{i_1j_2}+b_{i_0j_2}-b_{i_0j_3}+b_{i_1j_3}-b_{i_1j_2}+b_{i_2j_2}-b_{i_2j_1}+b_{i_0j_1}-b_{i_0j_0}=-(b_{i_0j_0}-b_{i_1j_0})+(b_{i_0j_1}-b_{i_2j_1})-(b_{i_1j_2}-b_{i_2j_2})+(b_{i_0j_2}-b_{i_1j_2})-(b_{i_0j_3}-b_{i_1j_3})=-D_{ij_0}+D_{i'j_1}-D_{i''j_2}+D_{ij_2}-D_{ij_3}$. 

As a whole, if $-D_{ij_0}+D_{i'j_1}-D_{i''j_2}+DD_{ij}=0$ then the Tanner graph has 10-cycles. In this item if Fossorier's equation is used then $432(_{3}^{m})(_{4}^{n})$ equations are required to investigate.
\item Take $j_0,j_1,j_2,j_3,j_4$ as five columns of a $3\times5$ submatrix. For this item there are $3\times6(_{3}^{5})=180$ cases to consider. Because, there are $(_{3}^{5})$ choices for obtaining $3\times3$ submatrices. For each of them there are 6 considerations related to 6-cycles. The $2\times2$ submatrices occur in the other two columns, so, the number of them are $(_{2}^{2})\times(_{2}^{3})=3$. The left side of Equation (1) related to one of them is presented in the following.

$b_{i_1j_0}-b_{i_1j_2}+b_{i_0j_2}-b_{i_0j_3}+b_{i_1j_3}-b_{i_1j_4}+b_{i_2j_4}-b_{i_2j_1}+b_{i_0j_1}-b_{i_0j_0}=-(b_{i_0j_0}-b_{i_1j_0})+(b_{i_0j_1}-b_{i_2j_1})-(b_{i_1j_4}-b_{i_2j_4})+(b_{i_0j_2}-b_{i_1j_2})-(b_{i_0j_3}-b_{i_1j_3})=-D_{ij_0}+D_{i'j_1}-D_{ij_4}+D_{i''j_2}+D_{ij_3}$. 

As a whole, if $-D_{ij_0}+D_{i'j_1}-D_{i''j_4}+DD_{ij}=0$ then the Tanner graph has 10-cycle. In this item if Fossorier's equation is used then $180(_{3}^{m})(_{5}^{n})$ equations are required to investigate.  

To prove the first condition of the theorem we utilise three items of case 1. If $3\times3$, $3\times4$ and $3\times5$ submatrices for 10-cycles are under consideration then we define two sets $\mathcal{A}$ and $\mathcal{B}$ for each $3\times3$ submatrix as follows. The set $\mathcal{A}$ includes the values obtained by the left side of equations in 6-cycle considerations. The set $\mathcal{B}$ includes all elements of three rows of the matrix, $DD$ equivalent to three rows of the exponent matrix under consideration. If $\mathcal{A}\cap \mathcal{B}\neq\emptyset$ then the Tanner graph has 10-cycles. Making use of such a manner significantly decreases the search space and all of values obtained in 6-cycle considerations are utilized which reduces the considerations in a great extent.
  
 {\bf Case 2}: Submatrices which include four rows of the exponent matrix. Suppose they are $i_0,\ i_1,\ i_2$ and $i_3$.

\item  Suppose three columns are $j_0,j_1,j_2$. The $4\times3$ submatrices of the exponent matrix are equivalent to $4\times3$ submatrices, $D'$, of the difference matrix, $D$. Like the previous cases, three elements of $D'$ which are on a distributed diagonal of a $3\times3$ submatrix of $D'$ are used to replace 6 terms of the left side of inequalities. There are $(_{3}^{4})$ choices for $3\times3$ submatrices of $D'$. The other row of $D'$ contains two elements which occur in the left side of Equation (1), when is rearranged by the elements of $D$. In the rearranged expression there is the sign "-" between two elements of $D'$. They occur in a row different from the three chosen rows for the $3\times3$ submatrix of $D'$, for which there are $(_{2}^{3})$ choices. So, they construct an element of the matrix $DD$. For this item there are $6\times4\times3=72$ equations to consider 10-cycles. And in the $m\times n$ exponent matrix there are $72\times{m\choose 4}{n\choose 3}$ equations to investigate 10-cycles.

The left side of one of Equations (1) and its corresponding equation whose elements belong to $D$ is as follows:

$b_{i_0j_0}-b_{i_0j_1}+b_{i_1j_1}-b_{i_1j_2}+b_{i_2j_2}-b_{i_2j_1}+b_{i_3j_1}-b_{i_3j_2}+b_{i_2j_2}-b_{i_2j_0}=(b_{i_0j_0}-b_{i_2j_0})-(b_{i_0j_1}-b_{i_1j_1})-(b_{i_1j_2}-b_{i_2j_2})-(b_{i_2j_1}-b_{i_3j_1})+(b_{i_2j_2}-b_{i_3j_2})=D_{i'j_0}-D_{ij_1}-D_{i''j_2}-D_{i'''j_1}+D_{i'''j_2}$.

As we see, the last two terms of the equation above, in the right side, is equivalent to one element of the matrix, $DD$. If it appears in $j$-th column of $DD$ then the equation $-D_{ij_1}+D_{i'j_0}-D_{i''j_2}-DD_{i'''j}=0$ provides 10-cycles. 

\item  Take a $4\times4$ submatrix of the exponent matrix, where $j_0,\ j_1,\ j_2$ and $j_3$  are their four columns. Like the previous item it is equivalent to a $4\times4$ submatrix of the difference matrix, $D$, we denote it by $D'$. We first choose a $3\times3$ submatrix of $D'$. In $D'$, there are a row and a column different from those in the chosen $3\times3$ submatrix which in their cross point contain an element appearing in equation based on the elements of $D'$. One of the other rows contains another element which occurs in the equation. For this element there are 9 possibilities to choose. So for each $4\times4$ submatrix of the difference matrix, $D$, there are $9\times6(_{3}^{4})(_{3}^{4})=864$ equations to consider. And as a whole, the number of 10-cycle considerations for this item is $864\times{m\choose 4}{n\choose 4}$.    
The left side of two of Equations (1) and their corresponding equations whose elements belong to $D$ are as follows:

$1)\ b_{i_0j_0}-b_{i_0j_1}+b_{i_1j_1}-b_{i_1j_2}+b_{i_2j_2}-b_{i_2j_3}+b_{i_3j_3}-b_{i_3j_2}+b_{i_1j_2}-b_{i_1j_0}=-(b_{i_0j_1}-b_{i_1j_1})+(b_{i_1j_2}-b_{i_3j_2})-(b_{i_2j_3}-b_{i_3j_3})+(b_{i_0j_0}-b_{i_1j_0})-(b_{i_1j_2}-b_{i_2j_2})=-D_{ij_1}+D_{i''j_2}-D_{i'''j_3}+D_{ij_0}-D_{i'j_2}$,

$2)\ b_{i_0j_0}-b_{i_0j_1}+b_{i_1j_1}-b_{i_1j_3}+b_{i_3j_3}-b_{i_3j_2}+b_{i_1j_2}-b_{i_1j_3}+b_{i_2j_3}-b_{i_2j_0}=-(b_{i_0j_1}-b_{i_1j_1})+(b_{i_0j_0}-b_{i_2j_0})-(b_{i_1j_3}-b_{i_3j_3})-(b_{i_1j_3}-b_{i_2j_3})+(b_{i_1j_2}-b_{i_3j_2})=-D_{i'j_1}+D_{ij_0}-D_{i''j_3}-D_{i'''j_3}+D_{i'''j_2}$.

As we see, in the first case the last two terms do not occur in the same row but for the second case they do. Therefore, in the second case we replace these two terms by one element of the matrix, $DD$.  
\item  Take a $4\times5$ submatrix of the exponent matrix, where $j_0,\ j_1,\ j_2,\ j_3$ and $j_4$  are their four columns. Like the previous item it is equivalent to a $4\times5$ submatrix of the difference matrix, $D$, we denote it by $D'$. We first choose a $3\times3$ submatrix of $D'$. In $D'$, there are a row and two columns different from those in the chosen $3\times3$ submatrix which in their cross point contain two elements appearing in equation based on the elements of $D'$. For these elements there are 4 possibilities to choose. So for each $4\times5$ submatrix of the difference matrix, $D$, there are $4\times6(_{3}^{4})(_{3}^{5})=960$ equations to consider. And as whole, the number of 10-cycle considerations for this item is $960\times{m\choose 4}{n\choose 4}$.

The left side of one of Equations (1) and its corresponding equation whose elements belong to $D$ is as follows:

$ b_{i_1j_0}-b_{i_1j_2}+b_{i_3j_2}-b_{i_3j_3}+b_{i_2j_3}-b_{i_2j_4}+b_{i_3j_4}-b_{i_3j_1}+b_{i_0j_1}-b_{i_0j_0}=-(b_{i_0j_0}-b_{i_1j_0})+(b_{i_0j_1}-b_{i_3j_1})-(b_{i_1j_2}-b_{i_3j_2})+(b_{i_2j_3}-b_{i_3j_3})-(b_{i_2j_4}-b_{i_3j_4})=-D_{ij_0}+D_{i'j_1}-D_{i''j_2}+D_{i'''j_3}-D_{i'''j_4}$.
As we see, the last two terms are an element of the matrix, $DD$. So, if it occurs in $k$-th row and $j$-th column of $DD$ then the equality  $-D_{ij_0}+D_{i'j_1}-D_{i''j_2}+DD_{i'''j}=0$ provides 10-cycles.

 {\bf Case 3}: Submatrices which include five rows of the exponent matrix. Suppose they are $i_0,\ i_1,\ i_2,\ i_3$ and $i_4$.

\item  Suppose three columns are $j_0,j_1,j_2$. We provide the left side of the two Equations (1) and their corresponding equations whose elements belong to $D$ in order to show that this item is similar to the item 5. So this item is omitted which declines the search space significantly. 

$1)\ b_{i_0j_0}-b_{i_0j_1}+b_{i_1j_1}-b_{i_1j_2}+b_{i_2j_2}-b_{i_2j_1}+b_{i_3j_1}-b_{i_3j_2}+b_{i_4j_2}-b_{i_4j_0}=(b_{i_0j_0}-b_{i_4j_0})-(b_{i_0j_1}-b_{i_3j_1})-(b_{i_3j_2}-b_{i_4j_2})+(b_{i_1j_1}-b_{i_2j_1})-(b_{i_1j_2}-b_{i_2j_2})=D_{ij_0}-D_{i'j_1}-D_{i''j_2}-D_{i'''j_1}-D_{i'''j_2}$.

$2)\ b_{i_0j_0}-b_{i_0j_1}+b_{i_2j_1}-b_{i_2j_2}+b_{i_3j_2}-b_{i_3j_1}+b_{i_4j_1}-b_{i_4j_2}+b_{i_1j_2}-b_{i_1j_0}=(b_{i_0j_0}-b_{i_1j_0})-(b_{i_0j_1}-b_{i_2j_1})-(b_{i_2j_2}-b_{i_3j_2})+(b_{i_1j_2}-b_{i_4j_2})-(b_{i_3j_1}-b_{i_4j_1})=D_{ij_0}-D_{i'j_1}-D_{i''j_2}+D_{i'''j_1}-D_{i''''j_2}$.

\item  Take $5\times4$ submatrices of the exponent matrix, where $j_0,\ j_1,\ j_2$ and $j_3$  are their four columns. This item is also similar to item 5 and is omitted from the search.

To prove the second condition of the theorem we make use of items 4 to 8. According to the examples provided in the mentioned items we result the condition.

\item  If five columns are $j_0,\ j_1,\ j_2,\ j_3,\ j_4$. In this item all elements of a $5\times5$ submatrix of $D$ are considered. If for every five elements which occur in the five rows and the five columns the obtained expression from rearranging the left side of Equation (1) is non-zero modular $N$ then such submatrices do not provide 10-cycles. From this item we conclude the third condition of the theorem.
\end{enumerate}
\end{prof}
\end{document}